\documentclass[prb]{revtex4-1}

\usepackage[english]{babel}
\usepackage[utf8]{inputenc}

\usepackage{graphicx}
\usepackage[utf8]{inputenc}

\usepackage{amsmath, amsfonts, amssymb, mathrsfs} 
\usepackage{braket}

\usepackage{xcolor}
\usepackage{xargs}
\usepackage{changebar}
\usepackage[colorinlistoftodos,prependcaption,textsize=tiny]{todonotes}
\newcommandx{\kzcomments}[2][1=]{\todo[linecolor=cyan,backgroundcolor=cyan!25,bordercolor=cyan,#1]{#2}}
\newcommandx{\unsure}[2][1=]{\todo[linecolor=red,backgroundcolor=red!25,bordercolor=red,#1]{#2}}
\newcommandx{\change}[2][1=]{\todo[linecolor=blue,backgroundcolor=blue!25,bordercolor=blue,#1]{#2}}
\newcommandx{\info}[2][1=]{\todo[linecolor=green,backgroundcolor=green!25,bordercolor=OliveGreen,#1]{#2}}
\newcommandx{\improvement}[2][1=]{\todo[linecolor=magenta,backgroundcolor=magenta!25,bordercolor=magenta,#1]{#2}}
\newcommandx{\thiswillnotshow}[2][1=]{\todo[disable,#1]{#2}}

\newcommand{\refeq}[1]{\!(\ref{#1})}

\newcommand{\adOp}{\hat{a}^\dagger}
\newcommand{\aOp}{\hat{a}}
\newcommand{\Hop}{\hat{H}}

\newcommand{\Nop}{\hat{n}}
\newcommand{\Cop}{\hat{c}}
\newcommand{\Cdop}{\hat{c}^\dagger}

\newcommand{\upOp}[1]{\Cop_{#1 \up}^\dagger} 
\newcommand{\downOp}[1]{\Cop_{#1 \down}^\dagger} 
\newcommand{\updownOp}[1]{\Cop_{#1 \up}^\dagger \Cop_{#1 \down}^\dagger} 

\newcommand{\vacuo}{\ket{0}}

\newcommand{\emptys}{\!\rule{.5em}{.4pt}\!}
\newcommand{\up}{\uparrow} 
\newcommand{\down}{\downarrow} 
\newcommand{\updown}{\uparrow \hspace{-0.225em}\downarrow} 

\newcommand{\Trace}{\text{Tr}}
\newcommand{\Trj}[2]{\text{Tr}_{#1} \left[ #2 \right]}

\newcommand{\expectv}[1]{\langle #1 \rangle}

\newcommand\NEW[1]{{\color{black} #1}}

\usepackage[]{tikz}
\usetikzlibrary{calc, shapes}

\begin{document} 

\title{Entanglement in finite quantum systems under twisted boundary conditions} 

\author{Krissia Zawadzki}
\affiliation{Departamento de Física e Ciência Interdisciplinar, Instituto de Física de São Carlos,
  University of São Paulo, Caixa Postal 369, 13560-970 São Carlos, SP, Brazil} \author{Irene
  D'Amico} \affiliation{Department of Physics, University of York, York, YO10\,5DD, United Kingdom}
\affiliation{Departamento de Física e Ciência Interdisciplinar, Instituto de Física de São Carlos,
  University of São Paulo, Caixa Postal 369, 13560-970 São Carlos, SP, Brazil}\author{Luiz
  N. Oliveira} \affiliation{Departamento de Física e Ciência Interdisciplinar, Instituto de Física
  de São Carlos, University of São Paulo, Caixa Postal 369, 13560-970 São Carlos, SP, Brazil}
  
\begin{abstract}
In a recent publication, we have discussed the effects of 
boundary conditions in finite quantum systems and their connection with symmetries. 
Focusing on the one-dimensional Hubbard Hamiltonian under twisted boundary conditions,  
we have shown that properties, such as the ground-state and gap energies, 
converge faster to the thermodynamical limit ($L \rightarrow \infty$) if a special torsion $\Theta^*$ is 
adjusted to ensure particle-hole symmetry. 
Complementary to the previous research, the present paper extends our analysis to 
a key quantity for understanding correlations in many-body systems: the entanglement. 
Specifically, we investigate the average single-site entanglement $\expectv{S_j}$ as a function 
of the coupling $U/t$ in Hubbard chains with up to $L=8$ sites 
and further examine the dependence of the per-site ground-state $\epsilon_0$ on the torsion $\Theta$ in different coupling regimes. 
We discuss the scaling of $\epsilon_0$ and $\expectv{S_j}$ under $\Theta^*$ and analyse their convergence 
to Bethe Ansatz solution of the infinite Hubbard Hamiltonian. 
Additionally, we describe the exact diagonalization procedure used in our numerical calculations 
and show analytical calculations for the case-study of a trimer.
\end{abstract}

\maketitle

\section{\label{sec:intro} Introduction}

The study of many-body phenomena has gained a new perspective with the recent  
collaboration between condensed matter (CM) and quantum information theory (QIT).  
Experimentally, technical advances fostered by QIT have allowed 
for a high control of nanoscale set-ups, turning into reality the possibility to simulate condensed
matter models \cite{Jordens-Nat.455.204,Byrnes-PRB.78.075320,Salfi-Nat-11342,Baier-Science-352.6282} 
and to measure their properties with single
site resolution \cite{Edge-PRA.92.063406,Parsons-Sci.353.1253,Boll-Sci.353.1257}.
At a fundamental level, both communities have brought contributions to 
our understanding of quantum correlations. The concept of entanglement has become 
a key ingredient to investigate collective 
behavior arising from microscopic degrees of freedom, such as critical properties and quantum phase transitions 
\cite{Vidal-PRL.90.227902,Amico-RMP.80.517,Laflorencie-PhysRep.646.0370}. 
In this context, a problem which have been receiving special attention
deals with the conditions under which properties in the thermodynamical limit can be accurately assessed 
by means of finite systems \cite{Gammel-SM.57.4437}, in which boundary conditions play a crucial role in the system's symmetries. 
From the experimental point of view, this issue is equivalent to attenuating  
finite-size effects by means of a given quantum protocol \cite{Papanastasiou-PRA.96.042332,Kormos-NatPhys.13.3,wang2018floquet}; analytically, 
it is an important ingredient for improving numerical methods for many-body systems \cite{Kent-PRB.59.1917}, 
such as exact diagonalization \cite{Gammel-SM.57.4437,Gros-PRB.53.6865}, Monte Carlo Simulations 
\cite{Lin-PRE.64.016702,Dagrada-PRB.94.245108} 
and Renormalization-Groups \cite{Cirac-AP.387.100,Yang-PRL.118.110504}.

In a previous work \cite{Zawadzki-BJP.47.5}, we have examined this question by addressing the 
compatibility between boundary conditions and conserved quantities in the finite one dimensional Hubbard Hamiltonian 
\cite{Hubbard-PRoyalSL.276.1365238,Lieb-PA.321.01}. We analysed some properties of small Hubbard chains   
under open, periodic and twisted boundary conditions and presented results 
for the ground state and gap energies, and local densities and magnetizations at half-filling. 
Special attention was given to the case of twisted boundary conditions \cite{Shastry-PRL.65234}, \NEW{
in which the ends of the chain are connected with a hopping amplitude having a torsion phase $\Theta$. }
\NEW{This situation is physically equivalent to a Hubbard chain coupled to an external vector potential $A$.
The effect of twisted boundary conditions in integrable models has been widely discussed \cite{Alcaraz-Annal-182.2,Shastry-PRL.65234,Shiroishi-JPSJ.66.8,Yue-JPA-30.3.849,Fukui-PRB.58.16051}. 
A vast literature on the low-energy properties covering analytical calculations via Bethe Ansatz of the ground-state energies, correlation functions and order parameters in spin chains, including the Hubbard Hamiltonian \cite{Shastry-PRL.65234}. In particular, it has been demonstrated that properties of Hamiltonians obeying U(1) symmetry do not deppend on the boundary condition as the system size is increased, the dependence being exponentially suppressed \cite{Shastry-PRL.65234}. 
Our motivation to revisit the case of finite and relatively small chains has its origins in recent studies of qubit systems.  The discussion is linked especially to the experimental realization of protocols engineering few particle systems, which, in practice, are no longer described by integrable models \cite{Rigol-PRL.103.100403,Weinberg-PR.688.1}. }
\NEW{In an attempt to be pedagogical,} we demonstrated that there is a special torsion $\Theta^* = \pi L/2$, where $L$ is the system size, in which one can preserve most symmetries of the infinite Hubbard model. 
\NEW{We argued that} an important consequence of this finding is that under twisted boundary conditions with $\Theta^*$, 
properties converge faster to the limit $L \rightarrow \infty$, with excellent results already 
for relatively small chains of, up to $7$ sites.

Here, we complement our analysis of the twisted boundary condition targetting the special property 
of entanglement. \NEW{We also provide an instructive description of a flexible numerical procedure 
based on a binary approach for small spin lattices, used to obtain the results in Ref. \cite{Zawadzki-BJP.47.5} 
and in the present paper.} \NEW{Here, we report numerical calculations for the ground-state energy and 
the average single-site entanglement,  
exploring their deviations from the thermodynamical limit ($L \rightarrow \infty$) as a function of the torsion $\Theta$ and the re-scaled coupling $U/t$.} \NEW{Equivalent results could be obtained from the Bethe Ansatz solution of the finite Hubbard Hamiltonian 
under twisted boundary conditions \cite{Bannister-PRB.61.4651}.}
Our results indicate that the 
entanglement under $\Theta^*$ converges fast, like 
the ground-state energy. 
We identify different behaviors in chains with odd and even number of sites  
\NEW{and their scaling as $L$ is increased. }

\NEW{The present paper is organized as follows. }In section \ref{sec:Hubbard} 
we review the connection between symmetries and 
boundary conditions. In section \ref{sec:numerical}, we explain the numerical procedure devised 
to perform the exact diagonalization of the many-body Hubbard Hamiltonian. 
Numerical results for the per-site ground-state energy and single-site entanglement  
of half-filled chains are presented in section \ref{sec:results}. Finally, we conclude our analysis on the 
scaling properties of Hubbard chains under twisted boundary conditions in section \ref{sec:conclusion}.
In appendix \ref{apx:trimer}, we also include the case-study of the Hubbard trimer, calculating explicitly 
the matrix Hamiltonian under twisted boundary condition and presenting analytical 
results for the single-site entanglement. 
\NEW{We show additional results for the deviations from the limit $L\rightarrow \infty$ in the single-site entanglement as a function of the torsion $\Theta$ and the coupling $U/t$ in appendix \ref{apx:Sj_Theta}.
}

\section{\label{sec:Hubbard} Symmetries of the Hubbard Hamiltonian under twisted boundary conditions}

The Hubbard Hamiltonian is one of the most studied models in condensed matter physics. 
It has grounded most of our knowledge of a wide class of solid state systems, ranging 
from conductors to insulators. More recently, it has been successfully used to investigate    
exotic states of matter occuring in 
quantum dots \cite{Byrnes-PRB.78.075320}, ultracold fermionic atoms and ion traps \cite{Salfi-Nat-11342,Jordens-Nat.455.204}, 
Bose-Einstein condensates \cite{Baier-Science-352.6282}, etc.

Comprising two terms - namely, 
the hopping and the Coulombian interaction - 
it translates in a simple way the competition between the localization and de-localization trends of quantum particles 
in a lattice. In one dimension, the fermionic Hubbard Hamiltonian including boundary conditions is written as 
\begin{align}
	\label{eq:Hubbard_BC}
	\hat{H} & =  - t \sum_{\ell = 1}^{L-1} (\Cop_{\ell+1}^\dagger \Cop_{\ell} + H.c.) 
			+ U \sum_{\ell=1}^{L} \Nop_{\ell \up} \Nop_{\ell \down} 
			\nonumber \\
			& \hphantom{=}
			- \mu \sum_{\ell=1}^{L} \Nop_{\ell}
			+ \hat{H}_{\text{BC}} ,
\end{align} 
where the operators $\Cop_\ell (c^\dagger_\ell)$ anihilates (creates) an electron at site $\ell$, 
$\Nop_{\ell, \sigma} = \Cop^\dagger_{\ell, \sigma} \Cop_{\ell, \sigma}$ counts the occupation of electrons with 
spin $\sigma$ at site $\ell$, $t$ is the hopping amplitude, $U$ accounts for the Coulomb repulsion penalyzing 
double occupation and $\mu$ is the chemical potential. The last term connects the ends of 
the chain $\ell = 1$ and $\ell = L$ and defines the boundary condition (BC)
\begin{align}
	\label{eq:Hboundary}
	\hat{H}_{\text{BC}} & = - ( \tau \Cop_{1}^\dagger \Cop_{L} + \tau^{*} \Cop^\dagger_L \Cop_1 ),
\end{align}
so that 
\begin{align}
	\label{eq:boundary_conditions}
	\tau & = \begin{cases}
	        0 & \text{open (OBC)} \\
	        t & \text{periodic (PBC)}\\
	        t e^{i\Theta} & \text{twisted (TBC)} 
	        \end{cases},
\end{align}
$0 < \Theta < \pi$being the torsion phase. \NEW{\footnote{We note that phases $0 \leq \Theta \leq \pi$ are symmetric to those from $ 2\pi$ to $\pi$ 
with the first half of trigonometric circle by 
$\Theta \rightarrow 2\pi - \Theta$.}}

\begin{figure}
	\begin{center}
		\includegraphics[]{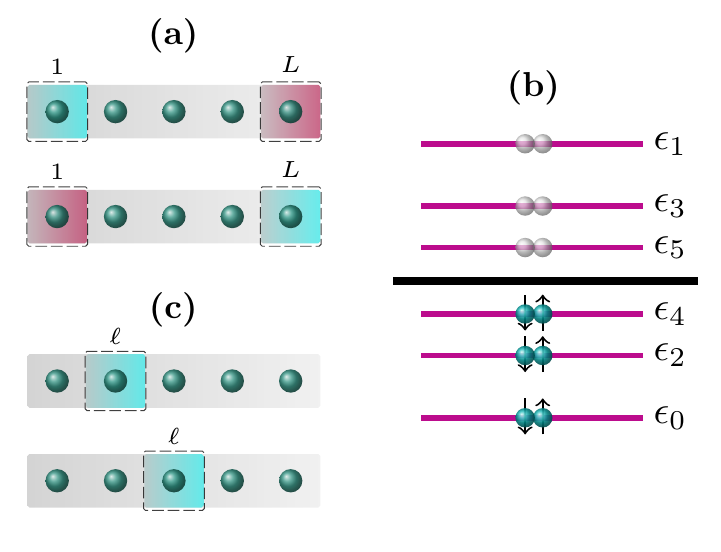}
	\end{center}
	
	\caption{\label{fig:symmetries} Some of symmetries present in the infinite one-dimensional Hubbard Hamiltonian. 
	\textbf{(a)} inversion symmetry: re-labelling sites from left to right and vice-versa does not change the transformed Hamiltonian.
	\textbf{(b)} particle-hole symmetry: the energetic cost to add a particle to the first unoccupied level is the same of removing a particle 
	at the last occupied level.
	\textbf{(c)} translation symmetry: the system remains invariant by shifting sites to its neighbors, so that 
	the linear momentum is conserved.}
\end{figure}

	Increasing $L \rightarrow \infty$ and keeping the average electron density constant 
	we recover the thermodynamical limit and the Hamiltonian becomes independent of 
	the boundary condition. \NEW{This limit is particularly interesting due to the variety of symmetries of the infinite Hubbard chain: 
	besides conserving charge, spin and spin rotation, which are symmetries present also in finite chains under any of the boundary conditions 
	defined in eq. \refeq{eq:boundary_conditions}, 
	the infinite model also possesses inversion, translation and particle-hole symmetry; the 
	latter having a crucial role.} An illustration of the last three is 
	shown in figure \ref{fig:symmetries} and a complete discussion can be 
	found in ref. \cite{Zawadzki-BJP.47.5}. 
    	\NEW{Here, we recapitulate, in more detail, the derivation of} the condition that the twist phase $\Theta$ must fulfill 
    	to preserve particle-hole and 
	translation symmetries 
	in a finite Hubbard chain under twisted boundary condition.

	In the presence of particle-hole symmetry, the cost for adding or removing a particle from the Fermi level is the same.
	The Hamiltonian must be invariant under the transformation
	\begin{align}
		\label{eq:phs}
		\Cop_\ell \rightarrow (-1)^\ell \adOp_\ell.
	\end{align} 

	Carrying out the transformation for the Hamiltonian under open boundary condition $\hat{H}_{OBC}$, we find 
	\begin{align}
		\label{eq:phs_OBC}
		 \hat{H}_{OBC} & =- t \sum_{\ell = 1}^{L-1} 
		 (-1)^{2\ell+1} (
		  \aOp_{\ell+1} \adOp_{\ell}
		  +
		  \aOp_{\ell} \adOp_{\ell+1}
		 )
			+ U \sum_{\ell=1}^{L} (-1)^{4\ell} \aOp_{\ell \up} \adOp_{\ell \up} \aOp_{\ell \down} \adOp_{\ell \down} 
			- \mu \sum_{\ell=1}^{L} (-1)^{2\ell} (\aOp_{\ell \up} \adOp_{\ell \up} + \aOp_{\ell \down} \adOp_{\ell \down} )
	\nonumber \\
	& = - t \sum_{\ell = 1}^{L-1} 
		 (
		  \adOp_{\ell} \aOp_{\ell+1} 
		  +
		  \adOp_{\ell+1} \aOp_{\ell} 
		 )
			+ U \sum_{\ell=1}^{L} \adOp_{\ell \up} \aOp_{\ell \up} \adOp_{\ell \down} \aOp_{\ell \down}
			- \mu \sum_{\ell=1}^{L} (\adOp_{\ell \up} \aOp_{\ell \up} + \adOp_{\ell \down} \aOp_{\ell \down} ),
	\end{align}
so that the open chain remains invariant.

	Carrying out the transformation for $\Hop_{BC}$, we obtain 
\begin{align}
	\label{eq:phs_TBC}
	\hat{H}_{\text{BC}} & = - 
	( \tau 
	(-1)^{1} \aOp_1 (-1)^{L} \adOp_L + 
	 \tau^{*} (-1)^{L} \aOp_L (-1)^1 \adOp_1
	 )
	 \nonumber \\
	 & = (-1)^{L+1} \tau \adOp_L \aOp_1
	 + (-1)^{L+1} \tau^{*} \adOp_1 \aOp_L 
	 .
\end{align}	

	To respect particle-hole symmetry, the hopping amplitude in the boundary 
	must fulfill the condition
	\begin{align}
		\label{eq:phs_TBC_condition}
		(-1)^{L+1} \tau^* & = - \tau.
	\end{align}

	As shown in Ref. \cite{Zawadzki-BJP.47.5}, condition \refeq{eq:phs_TBC_condition} is fullfilled if the torsion is 
	adjusted as follows
	\begin{align}
		\label{eq:phs_TBC_condition3}
		\Theta^* & = \frac{\pi L}{2}.
	\end{align}	
	
	For $0 \leq \Theta \leq \pi$, the previous relation split the cases of even and odd $L$'s 
	\begin{align}
		\label{eq:phs_condition_Loddeven}
		\Theta^{*}_{\text{odd}} & = \frac{\pi}{2}
		\nonumber \\
		\Theta^{*}_{\text{even}} & = 
		\begin{cases}
		0, & \text{$L/2$ even}
		\\
		 \pi, & \text{$L/2$ odd}
		\end{cases}
		.
	\end{align}
	
	Now, we can carry out a similar sequence of steps to find the conditions for the phase $\Theta$ so 
	that the system conserves momentum. Starting from the transformation 
	\begin{align}
		\label{eq:translation}
		\Cop_\ell \rightarrow e^{i\ell \theta} \aOp_\ell,
	\end{align}
	and inserting it into $\Hop$ only change the first and last terms of eq. \refeq{eq:Hubbard_BC}, giving 
	\begin{align}
		\label{eq:HBC_hopping}
		- t \sum_{\ell=1}^{L-1} (\Cop_{\ell+1}^\dagger \Cop_{\ell} + H.c.) 
		 & = 
		- \sum_{\ell=1}^{L-1} (e^{-i\theta} \aOp_{\ell+1}^\dagger \aOp_{\ell} + 
		e^{+i\theta} \aOp_{\ell+1}^\dagger \aOp_{\ell}),
	\end{align}
	and 
	\begin{align}
		\label{eq:HBC}
		\Hop_{BC}
		 & = 
		- t (e^{i(L-1)\theta +\Theta} \aOp_{1}^\dagger \aOp_{L} + 
		e^{-i(L-1)\theta-\Theta} \aOp_{L}^\dagger \aOp_{1}).
	\end{align}

	The invariance of the Hamiltonian under \refeq{eq:translation} requires 
	\begin{align}
		\label{eq:theta}
		\theta = \frac{\Theta}{L},
	\end{align}
		which is equivalent of implementing a local twist through the hopping amplitudes
	\begin{align}
		\label{eq:t_theta}
		t \rightarrow t e^{i\theta},
	\end{align}
	so that the torsion $\Theta$ is distributed along the chain.	
	
	These valuable findings bring the interesting question of how  
	properties in the thermodynamical limit compare with those in 
	finite chains when the boundary conditions fulfill conservation laws.
	Here, we are going to examine the dependencies of energies 
	and correlations with the torsion $\Theta$, discussing their 
	convergence to their values for the infinite Hubbard model calculated exactly 
	with the Bethe Ansatz.
	
	We now consider the distances $\Delta \epsilon_p = |\epsilon_p (L, n, U, \Theta)-\epsilon_p^{BA}(n,U)|$ \NEW{between the 
	Bethe Ansatz density energies $\epsilon_p^{BA}(U,n)$  for the infinite Hubbard model with filling $n$ 
	and } 
	\begin{align}
		\label{eq:persite_ergs}
		\epsilon_p(L, n, U, \Theta) & = 
		\frac{1}{t \, L}\bra{\Psi_p} \hat{H}(L, n, U, \Theta) \ket{\Psi_p},
	\end{align}
	is the per-site energy,
	where $\ket{\Psi_p}$ is the exact $p$-th state of $\hat{H}$ \NEW{($p=0$ corresponds to the ground-state, $p=1$ corresponds to the 
	first excited state and so on)} calculated for 
	a chain with $L$ sites, $Q = nL$ particles, coupling $U$ and torsion $\Theta$.
	
	In particular, the ground-state \NEW{density energy $\epsilon_0(L, n=1, U, \Theta)$ at half-filling ($n=1$ or $Q = L$) will be 
	compared with the Bethe Ansatz ground-state energy for the infinite half-filled Hubbard model, calculated from }
	\begin{align}
		\label{eq:eGSBA}
		\epsilon_0^{BA}(n=1, U) & = -4 \int_0^\infty dx \frac{J_0(x) J_1(x)}{x[1+e^{Ux/2}]}
		.
	\end{align}

	The single-site entanglement can be quantified by means of the Von Neumann entropy 
	\begin{align}
		\label{eq:Von_Neumann_S}
		S(\rho_\ell) &= -\sum_p \lambda_p \log(\lambda_p),
	\end{align}
	where $0 \leq \lambda_p \leq 1$ are the eigenvalues of the reduced density matrix 
	$
		\rho_\ell  = \Trj{k \neq \ell}{\rho}
	$, defined for pure states 
	$\rho = \ket{\Psi} \bra{\Psi}$ and 
	obtained by tracing all the degrees of freedom of the system \NEW{excluding the 
	degrees of freedom of site $\ell$.}

	For fermions, the reduced density matrix is a $4 \times 4$ diagonal matrix whose 
	eigenvalues correspond to 
	the probabilities to find the $\ell$-th site empty ($\lambda_{0}$), single 
	($\lambda_\up$ and $\lambda_\down$) or 
	double occupied ($\lambda_{\updown}$). Explicitly, it reads 
	\begin{align}
		\label{eq:SSreduced_density_matrix3}
		\rho_\ell & = 
		\begin{pmatrix}
		\lambda_{0} & 0 & 0 & 0\\ 
		0 & \lambda_{\down} & 0 & 0\\ 
		0 & 0 & \lambda_{\up} & 0\\ 
		0 & 0 & 0 & \lambda_{\updown}
		\end{pmatrix}.
	\end{align}

	In our comparision of the average single-site entanglement entropy $\expectv{S_j}(L, n, U, \Theta^*)$ with 
	the Bethe Ansatz results for the infinite Hubbard model, we will rely on the  
	analytical expression derived in \NEW{Refs. \cite{Gu-PRL.93.086402,Larsson-PRL.95.196406,Franca-PRL.100.070403}.}
	It reads 
\begin{equation}
	\label{eq:SSem_BA}
	\expectv{S_j}(n, U) = - 2 
	\Bigg( \frac{n}{2} - \frac{\partial \epsilon}{\partial U} \Bigg) \log_2 \Bigg[ \frac{n}{2} - \frac{\partial \epsilon}{\partial U}  \Bigg] 
	- \Bigg( 1 - n+ \frac{\partial \epsilon}{\partial U} \Bigg) \log_2 \Bigg[  1- n + \frac{\partial \epsilon}{\partial U} \Bigg]
	- \frac{\partial \epsilon}{\partial U}  \log_2 \Bigg[\frac{\partial \epsilon}{\partial U}  \Bigg] 
	,
\end{equation}
where $\epsilon = \epsilon^{BA}_0(n,U)$ is the per-site ground-state energy, 
which will be computed from the Bethe Ansatz at half-filling $n=1$ as indicated in eq. \refeq{eq:eGSBA}.

	The numerical procedure we used to obtain these properties in the case of finite lattices 
	with given torsion $\Theta$ is described in the next section.

\section{\label{sec:numerical} Exact diagonalization procedure}

		The complete Hilbert space of a fermionic Hubbard chain with $L$ sites comprises 
		$4^L$ states, as each site can be empty, singly occupied with $\sigma=\up, \down$ 
		or doubly occupied. In practice, carrying out the diagonalization of the full 
		matrix Hamiltonian is limited to few sites, of the order of ten. 
		The conservation of charge and spin allows us to bring the Hamiltonian \refeq{eq:Hubbard_BC} into a 
		block diagonal form, so that each block $H(Q,S)$ is associated with 
		a smaller Hilbert space $(Q,S)$ formed by states with definite charge $Q$ and spin $S$. 
		\NEW{For chain with $L \leq 9$ sites, the number of states in the largest 
		subspace $(Q,S)$ (8820 states) can still be handled exactly without high computational efforts. In practice, larger matrices $L > 10$ would require the use of special techniques for storing and 
		diagonalizing the matrix Hamiltonian $H(Q,S)$, i.e., a large  
		space in memory RAM and a longer time to perform the diagonalization, whose complexity order scales with $\mathcal{O}(n^3)$, where $n$ 
		is the dimension of the problem. }

		In the present work, we \NEW{implemented} an exact diagonalization procedure in which the fermionic 
		states are represented in a binary notation \cite{Lin-PRB.42.6561,Lin-CP.7.4.400}. 
		\NEW{The use of hashing tables, like the binary notation, to represent quantum states of spin models  
		is a convenient choice in exact diagonalization procedures, including the Lanczos algorithm \cite{Gagliano-PRB.34.1677,Lin-PRB.42.6561}. 
		The latter allows to obtain with high accuracy the low-energy spectrum of chains with up to $L = 24$ sites 
		\cite{Lin-CP.7.4.400} 
		for open boundary conditions and $L=12$ for twisted boundary conditions \cite{Gagliano-PRB.34.1677,Bannister-PRB.61.4651}, 
		the latter being halved because the twist phase introduces complex numbers in the Hamltonian matrix therefore requiring the double 
		of space and more computation time compared to real matrices. }
		
		\NEW{Here, we will describe a simple formulation of a binary hashing to represent quantum 
		states of one-dimensional fermionic systems, from which one is able to obtain exactly the full spectrum of 
		the Hamiltonian $H(Q,S)$ using a standard diagonalization routine. Differently from other 
		methods, such as the Lanczos diagonalization, the procedure yields all the $4^L$ eigenstates and eigenvalues 
		of the full many-body Hamiltonian without any approximation or truncation. The procedure comprises 
		three steps: the binary representation of basis elements in the subspace $(Q,S_z=S)$, 
		the rotation of the basis in $(Q,S_z)$ to the subspace $(Q,S)$, and the 
		projection of the Hamiltonian operator into the basis elements forming the subspace $(Q,S)$.}

\begin{figure}[h!]
\begin{center}
	\includegraphics[scale=1]{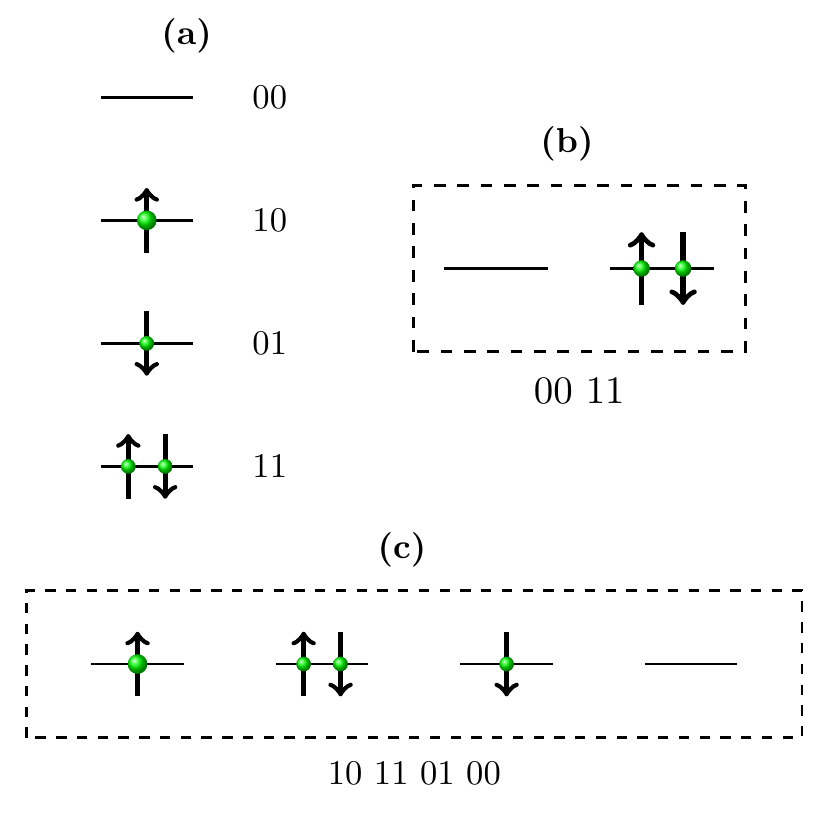}
\end{center}
\caption{\label{fig:bin_repr} Binary representation of fermionic configurations $\ket{b}$. 
Examples of configurations for \textbf{(a)} single-site with $i_b = 0, 2, 1, 3$ from top to bottom; 
\textbf{(b)} dimer $L = 2$ with $i_b = 3$;
and \textbf{(c)} tetramer $L = 4$ and $i_b = 180$.}
\end{figure}

		\NEW{The fist step of the exact calculation involves the definition of the hashing code to represent fermionic states in 
		the binary form.}
		Given a lattice of $L$ sites, each one of the 
		$n_B = 2^{2L}$ possible spin configurations $\ket{\vec{\sigma}} = \ket{\sigma_1} \otimes \ket{\sigma_2} \otimes ... \otimes \ket{\sigma_L}$		
		is associated with a sequence of $2L$ bits, the even indexes  
		referring to occupations of $\up$ and the odd indexes referring to occupations of $\down$, i.e.
		\begin{align}
			\label{eq:binconf}
			\ket{b} & = \ket{b_{1\up} \, b_{1\down} \,\, b_{2 \up} \, b_{2 \down} \, , ..., \,\, b_{L \up L} \, b_{L \down }}
			\nonumber \\
			& = \ket{\sigma_1} \otimes \ket{\sigma_2} \otimes ... \otimes \ket{\sigma_L}
			,
		\end{align}
		\NEW{where $b_{\ell \sigma}$ ($\ell=1,...,L$) can be either $0$ or $1$. $b_{\ell\up} = 1$ means that there is an electron $\up$ in the $\ell$-th site (it is equivalent to the creastion operator $c_{\ell\up}^\dagger$), whereas $b_{\ell\up} = 0$ indicates that site $\ell$ does not have an electron $\up$. Likewise, $b_{\ell\down} = 1$ is equivalent to $c_{\ell\down}^\dagger$ and $b_{\ell\down} = 0$ is equivalent to $c_{\ell\down}$.}
		 Some examples are illustrated in figure \ref{fig:bin_repr}.

	We can ascribe an integer label $i_b = 0, ..., n_B-1$ to each binary configuration $\ket{b}$, so that 
	\NEW{
	\begin{align}
		\label{eq:binary_to_index}
		i_b & = \sum_{\ell=1}^{L} ( b_{\ell \up} \, 2^{2L-\ell} \, + b_{\ell \down} \, 2^{2L-\ell-1}).
	\end{align} 
	}
	
	
	The label $i_b$ provides all information of the spin state $\sigma_\ell$ of each site $\ell$. 
	For instance, 
	\begin{align}
		\label{eq:bitUpj}
		b_{\ell\up} & = \text{mod}_2\left[\frac{i_b}{2^{2L-2\ell+1}}\right],
	\end{align}
	and 
	\begin{align}
		\label{eq:bitDmnj}
		b_{\ell\down} & = \text{mod}_2\left[\frac{i_b}{2^{2L-2\ell}}\right].
	\end{align}
	
	The action of any operator $\hat{O}$ on any state of the complete Hilbert space written in the binary form 
	of eq. \refeq{eq:binary_to_index} only requires the definition of such operator in terms of bit operations. 
	For instance, 
	the action of the creation and annihilation operators $\Cop_{\ell\up}^\dagger$ and $\Cop_{\ell\up}$ ($\ell=1,...,L$) 
	in a state $\ket{i_b}$ labeled by $i_b$ can be defined as
	\begin{align}
		\label{eq:example_binOp}
		\Cop_{\ell\up}^\dagger \ket{i_b} & \equiv \frac{1}{2}(1+(-1)^{b_{\ell\up}}) \ket{i_b+2^{2(L-\ell)+1}} \nonumber \\
		\Cop_{\ell\up} \ket{i_b} & \equiv \frac{1}{2}(1+(-1)^{1+b_{\ell\up}}) \ket{i_b+2^{2(L-\ell)+1}}.
	\end{align}		

	Similar relations can be defined for the $\down$ spins, replacing $2\ell \rightarrow 2\ell-1$ and adding an extra 
	factor $(-1)^{b_{\ell\up}}$ accounting for the fermionic signals.	
	
	As an example, consider the state $\ket{i_b = 9} = \ket{1 0 0 1}$ and the action of the operators 
	$\Cop_{1\down}^\dagger$ and $\Cop_{2\up}$. It follows that 
	\begin{align}
		\label{ex:a_ad}
		\Cop_{1\down}^\dagger \ket{i_b=9} & = \frac{1}{2} (1+(-1)^0) \ket{9 + 2^{2(2-1)}}
		= \ket{\tilde{i}_b = 13} 
		\nonumber \\
		\Cop_{2\up} \ket{i_b=9} & = \frac{1}{2} (1+(-1)^1) \ket{9 + 2^{2(2-2)+1}} = 0 \ket{\tilde{i}_b = 11}. 
	\end{align}
	
	Notice that while the operation $\Cop_{1\down}^\dagger \ket{i_b=9}$ yields a new binary index $\tilde{i}_b = 13$ which reconstructs 
	the binary state $\ket{1 1 0 1}$, the action of $\Cop_{2\up}$ on $ \ket{i_b=9}$ does not produce a new binary configuration, 
	as the spin $\up$ of the second site is empty.

	Once defined the operators composing the Hamiltonian in terms of binary relations, we are able to proceed 
	and construct the corresponding matrix $H$ in the subspaces of states with definite charge and spin. 	
	Within the binary approach, the subspaces $(Q,S)$ can be obtained in two steps. 
	First, we need to identify the $n_B$ configurations constrained by
		\begin{align}
			\label{eq:q_sum}
			\sum_{\ell = 1}^L b_{\up \ell} + b_{\down \ell} - L = Q,
		\end{align}
		\begin{align}
			\label{eq:sz_sum}
			\sum_{\ell = 1}^L b_{\up \ell} - b_{\down \ell} = S_z,
		\end{align}		
	which form the Hilbert space $(Q,S_z)$ of eigenstates of the operators $\hat{Q}$ and $\hat{S_z}$, defined as 
		\begin{align}
			\label{eq:Q_Op}
			\hat{Q} & = \sum_{\ell = 1}^{L}\sum_{\sigma=\up,\down} (\Cop_{\ell\sigma}^\dagger \Cop_{\ell\sigma} -\frac{1}{2}),
		\end{align}
		 and 
		 \begin{align}
			\label{eq:Sz_Op}
			\hat{S}_z & = \frac{1}{2} \sum_{\ell=1}^{L} \sum_{\mu, \nu = \up, \down} \Cop_{\ell\mu}^\dagger \sigma^{z}_{\mu, \nu}\Cop_{\ell\nu},
		\end{align}
\NEW{where $\sigma^z$ is the $z$ component of the Pauli matrices.}

	\NEW{The subspace $(Q,S_z)$ comprises $n_B$ binary configurations forming the basis $\{ \ket{q,s_z, b} \}$ $(b=0,1,...,n_B-1)$, 
	where 
	\begin{align}
		\label{eq:QSzeig}
		\hat{Q}\ket{q,s_z, b} & = Q \ket{q,s_z, b} \nonumber \\
		\hat{S}_z \ket{q,s_z, b} & = S_z \ket{q,s_z, b}
	\end{align}
	for all $b = 0, ..., n_B-1$.}
	
	\NEW{The conservation of the total spin $S$ yields a new projection of the Hamiltonian into 
	the subspace $(Q,S)$ comprising $n_P < n_B$ eigenstates $\{ \ket{q,s,s_z=s, p} \}$ $(p = 0, ..., n_P-1)$ of the operator 
		 \begin{align}
			\label{eq:S_Op}
			\hat{\vec{S}} & = \frac{1}{2} \sum_{\ell=1}^{L} \sum_{\mu, \nu = \up, \down} \Cop_{\ell\mu}^\dagger \vec{\sigma}_{\mu, \nu}\Cop_{\ell\nu},
		\end{align}
where $\vec{\sigma} = \sigma_x \hat{x} + \sigma_y \hat{y} + \sigma_z \hat{z}$.
	
	All $n_P$ basis elements in the subspace $(Q,S)$ satisfy
	\begin{align}
		\label{eq:QSzeig}
		\hat{\vec{S}}^2 \ket{q, s, s_z=s, p} & = s (s+1) \ket{q, s, s_z=s, p}.
	\end{align}	
	}

	Having the $n_B$ eigenstates $\{ \ket{q,s_z, b} \}$ and definitions for the operators in the binary notation, 
	\NEW{the projection of the Hamiltonian $\hat{H}$ into a basis definying the subspace $(Q,S)$ requires the transformation $\mathcal{T}_{p,b}$ 
	rotating the basis 
	$\{ \ket{q,s_z, b} \}$ in $(Q,S_z)$ into the new basis $\{ \ket{q,s,s_z=s, p} \}$ in $(Q,S)$.}
	This can be done using two approaches. We can define the operator $\hat{S}^2$ in 
	the binary form, project it into 
	the basis $\{ \ket{q,s_z, b} \}$ and from the diagonalization of the matrix $S^2$  
	identify the eigenstates with $s$ and $s_z=s$. Alternatively and, more efficiently, we 
	implement an iterative procedure in which the Hilbert spaces $(Q,S)$ are 
	constructed by growing the chain from $\ell = 1$ to $\ell=L$ and finding the 
	eigenstates 
	$\ket{q_\ell,s_\ell, s_{z\ell}, p_\ell}_\ell$ with help of the rules of 
	addition of angular momenta.

	We start with all eigenstates $\ket{q_{\ell=1},s_{\ell=1},s_z=s,p_{\ell=1}}$ for a single site in the end of the chain, which are 
	\begin{align}
		\label{eq:qs_single_site}
		\ket{q = 0, s = 0, s_z = 0, p = 0}_{\ell=1} & = \vacuo \nonumber \\
		\ket{q = 1, s = +\frac{1}{2}, s_z = +\frac{1}{2}, p = 0}_{\ell=1} & = \Cop^\dagger_{L\up} \vacuo \nonumber \\
		\ket{q = +2, s = 0, s_z = 0, p = 0}_{\ell=1} & =  \Cop^\dagger_{L\up} \Cop^\dagger_{L\down} \vacuo
		.
	\end{align}

\begin{figure}
\begin{center}
	\includegraphics[scale=1]{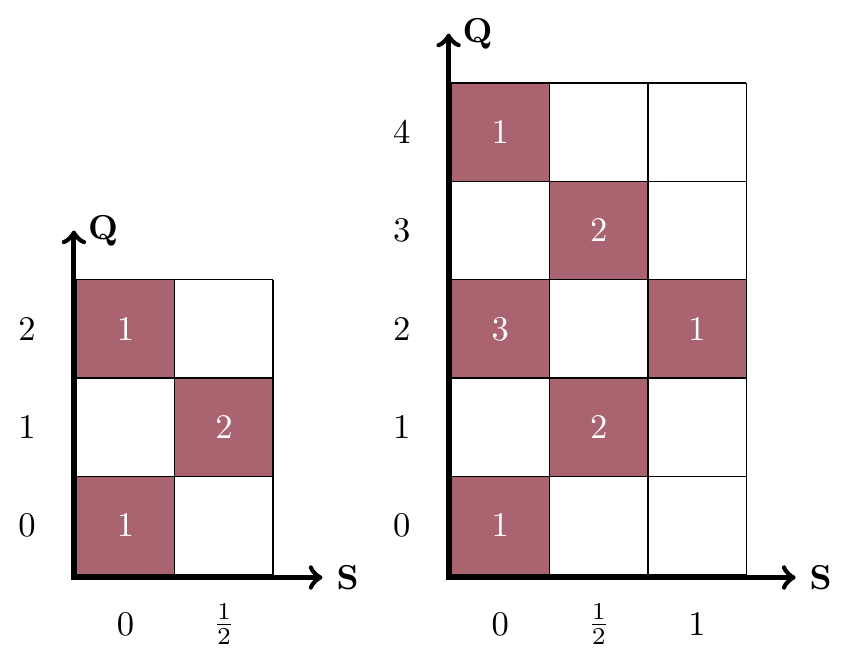}
\end{center}

\caption{\label{fig:QS_boards} Board of active sectors $(Q,S)$ in the iterations $\ell = 1$ and $\ell = 2$ and the number $p$
of states $\ket{q,s,s=s_z,p, \ell}$ with definite charge and spin inside them. 
Colored squares indicate the active sectors of the current chain.  White numbers mean the number of states in each sector.}
\end{figure}
	
	These states are stored in a board of active sectors, as illustrated in figure \ref{fig:QS_boards}. 
	The next step is adding the site $L-1$ with Clebsch-Gordan coefficients to obtain the eigenstates of 
	$\ket{q_{\ell=2},s_{\ell=2},s_z=s,p_{\ell=2}}$, which are stored in a new set of active sectors in 
	the board. For this, we use the following relations 
\NEW{
\begin{align}
	\label{eq:Osouth}
	\ket{q,s,s_z,p'}_{\ell+1} = \ket{q+1,s,s_z,p'}_{\ell} \quad p '= 0,.., n_P(q+1,s)_{\ell},
\end{align}

\begin{align}
	\label{eq:Oeast}
	\displaystyle
	\ket{q,s,s_z,p''+\max{p'}}_{\ell+1}
	& = c_{L-\ell\up}^\dagger \ket{q,s-\frac{1}{2},s_{z}-\frac{1}{2},p''}_{\ell} 
	\nonumber \\ 
	& p ''= 0,.., n_P(q,s-\frac{1}{2})_{\ell},
\end{align}	

\begin{align}
	\label{eq:Owest}
	\displaystyle 
	\ket{q,s,s_z,p'''+\max{p'}+\max{p''}}_{\ell+1} & = 
	\frac{-1}{\sqrt{2s + 1}} c_{L-\ell\up}^\dagger \ket{q,s+\frac{1}{2},s_{z}-\frac{1}{2},p'''}_{\ell} 
	\nonumber \\
		& 
	+ \sqrt{\frac{2s}{2s+1}} c_{L-\ell\down}^\dagger \ket{q,s+\frac{1}{2},s_{z}+\frac{1}{2},p'''}_{\ell}
	\nonumber \\ 
	& p'''= 0,.., n_P(q,s+\frac{1}{2})_{\ell},
\end{align}	

\begin{align}
	\label{eq:Onorth}
	\displaystyle
	\ket{q,s,s_z,p''''+\max{p'}+\max{p''}+\max{p'''}}_{\ell+1} & = c_{L-\ell\up}^\dagger c_{L-\ell\down}^\dagger \ket{q-1,s,s_z,p''''}_{\ell}
	\nonumber \\
	& p''''= 0,.., n_P(q-1,s)_{\ell}.
\end{align}	
}	
	Notice that, due to the degeneracy of the $z$ 
	components of momentum, fixing $s_z = s$ is convenient to save memory, so that we do not need 
	to store all the $s_z$ components of $s$ (states $\ket{q,s,s_z\neq +s, p}$) because they can be 
	simply recovered from $\ket{q,s,s_z=+s, p}$. For example, the the configurations $\ket{q,s, s_z = s -1, p}_{\ell}$ 
	needed to construct states $\ket{q,s\pm 1/2, s_z \pm 1/2, p}_{\ell+1}$
	from the ones obtained in the last iteration - $\ket{q,s, s_z = s, p}_{\ell}$ in eqs. \refeq{eq:Oeast} and \refeq{eq:Owest}- 
	can be implemented easily by flipping all the spins $\up$ of the previously stored $\ket{q,s, s_z = s, p}_{\ell}$.

	The growing procedure is repeated until $\ell = L$, yielding all the rotation matrices $\mathcal{T}_{p,b}(Q,S)$
	for each subspace $(Q,S)$ and their binary states $\ket{b}$. Using the bit rules for  
	the action of the operators defining the Hamiltonian $\hat{H}$, the matrix elements of $H(Q,S_z)$ are calculated as 
\begin{equation}
	H_{b,b'}  =\bra{{i_b}} \hat{H} \ket{{i_{b'}}},
\end{equation}
where ${i_b}$ and ${i_b'}$ label the binary configurations $\ket{b}$ and $\ket{b'}$ of $(Q,S)$.	
	
	The rotation $\mathcal{T}_{p,b}(Q,S)$ is then applied, resulting in a matrix 
	$n_P \times n_P$, i.e.,
	\begin{align}
		\label{eq:H_QS}
		H(Q,S) & = \mathcal{T}_{p,b}(Q,S) \, H(Q,S_z) \mathcal{T}^{-1}_{p,b}(Q,S)
		,
	\end{align}
	which can be diagonalized.
	
	In the table below, we present some examples of numbers $n_P$ and $n_B$ as a function of $L$.

\begin{table}[th!]

\begin{center}
	\begin{tabular}{c||c|c}
	\hline
	$L$  &  $n_P$ & $n_B$ \\
	\hline \hline
	2   & 	3	& 	4	\\
	3   & 	6	&  	9	\\
	4   & 	20	&   36		\\
	5   & 	75	&  	100	\\
	6   & 	175	&  	400	\\    
	7   & 	784	&  	1225	\\  
	8   & 	1764	& 4900 		\\  
	9   & 	8820	& 15876 		\\  
	10  &	19404	& 63504
	\end{tabular}
\end{center}
  
  \caption{\label{tab:1} Highest number $n_P$ of states in local Hilbert spaces $(Q,S)$ and number $n_B$ of 
  binary configurations needed to generate them as a function of $L$. For even $L$, the most dense Hilbert space 
  is $(Q=L,S=0)$, whereas for $L$ odd it is $(Q=L, S= \frac{1}{2})$.}
\end{table}

	Under twisted boundary conditions, once the matrix elements of $H$ are complex, effectively, 
	the memory needed to store the full Hamiltonian is $n_P^2$ double precision floating points, 
	which is twice the capacity needed for open and periodic boundary conditions.

	The procedure introduced above 
	was used in our numerical calculations of the ground-state 
	energy and single site entanglement presented in section \ref{sec:Hubbard}. Besides the 
	ground-state properties, our code provides the full excitation spectrum of $H(Q,S)$ for 
	any coupling $U$ and torsion $\Theta$. It also offers a flexible framework with support for non-homogeneous model parameters, non-local interactions and time-dependent calculations. 
	For the purposes of the present paper, we will 
	focus our analysis on the case of half-filled chains $Q = L$ and $S = 0$ ($L$ even) or 
	$S= 1/2$ ($L$ odd). Our results are presented below.

\section{\label{sec:results} Results}

\begin{figure}[htb!]
	
	\begin{center}
		\includegraphics[]{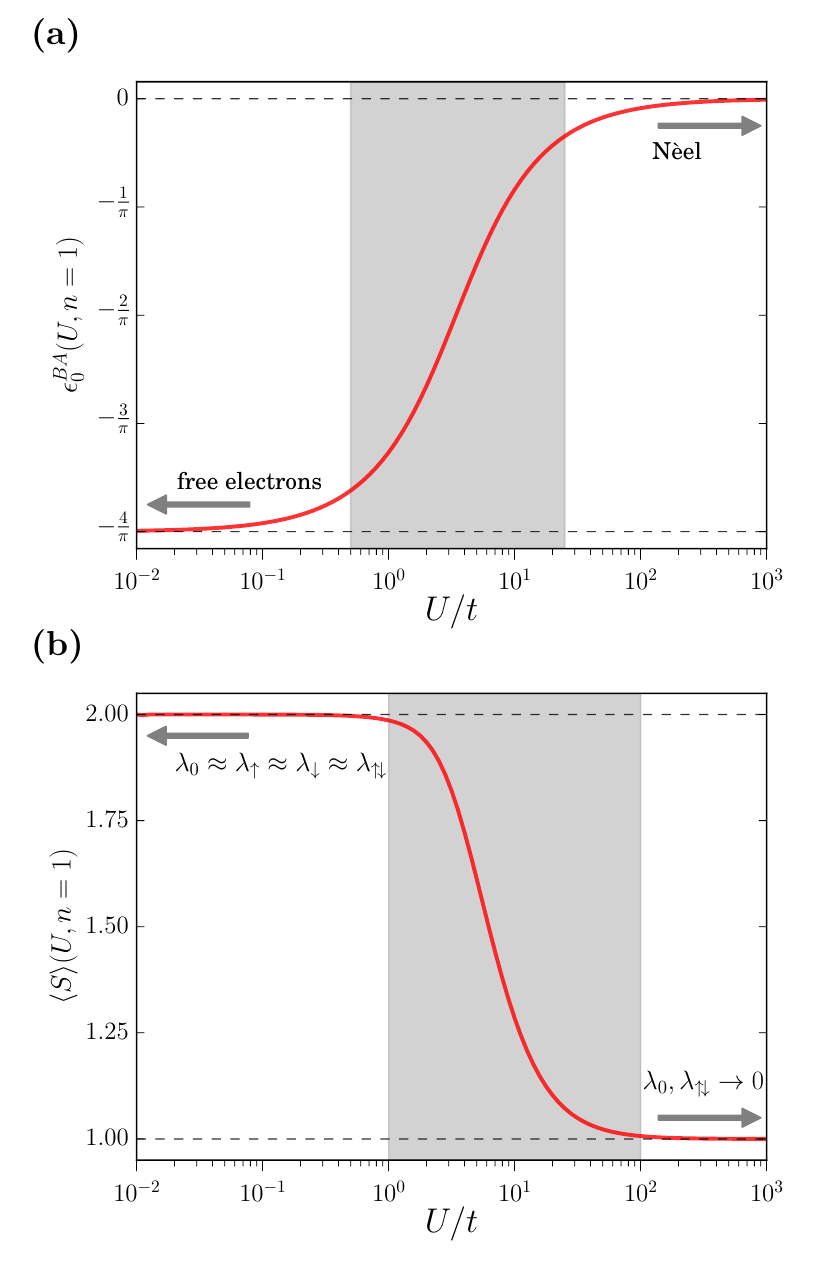}
	\end{center}
	
	\caption{\label{fig:BA_props} Ground-state energy \textbf{(a)} and single-site entanglement entropy 
	\textbf{(b)}
	for the infinite 1-D Hubbard model computed from the Bethe Ansatz solution at half-filling - 
	eqs. \refeq{eq:eGSBA} and \refeq{eq:SSem_BA}. The behavior of both quantities illustrates two solid state phases of 
	the Hubbard model: a tight-binding model for a free electron gas when the coupling is abscent ($U = 0$) 
	and a N\`eel antiferromagnetic insulator in the limit $U \rightarrow \infty$. In the shaded in region (gray), the Coulomb 
	repulsion is not negligible nor high enough, so that competing correlations 
	exist in the ground-state wave function.}
\end{figure}

	In the infinite chain, ground-state properties at half-filling 
	capture the rich physics regarding the phases of the Hubbard model. For $U = 0$, the Hamiltonian 
	$\Hop(L \rightarrow \infty)$ reduces to a free electron gas, as the electrons can move freely along the chain through the kinetic hopping. In the presence of non-zero coupling, 
	even infinitesimal, the system enters in an insulating phase, with gap energy 
	$\Delta$ increasing with $U/t$. In the limit $U \rightarrow \infty$, the prohibitive cost of double 
	occupation leads the system to become a N\`eel antiferromagnetic insulator. The change in the behavior of properties during 
	the transition from the non-interacting 
	($U/t \rightarrow 0$) to \NEW{extreme} Mott insulating ($U/t \gg 1$) phase is noticeable, 
	as illustrated in figure \ref{fig:BA_props}.  
	The per-site ground-state energy starts from its minimum value $\epsilon_0^{BA} (n=1,U \rightarrow 0) \approx - 4 /\pi$ 
	and increases up to zero, when electrons freeze in a state whose components only assume single occupation. See figure  \ref{fig:BA_props}\textbf{(a)}.

	The transition between these two extremes is particularly interesting 
	when we analyse internal correlations by means of the average single-site entanglement, shown in 
	figure \ref{fig:BA_props}\textbf{(b)}. When $U = 0$, the ground-state wave-function $\ket{\Psi_0}$
	decomposes in a Slater determinant of single-site orbitals with equal contribution. 
	Inspection of eq. \refeq{eq:SSreduced_density_matrix3} 
	provides a limiting value for $\expectv{S_j}(U \rightarrow 0) \rightarrow 2$, when all the eigenvalues of the 
	reduced density matrix are degenerate, i.e., 
	$\lambda_{0} \approx \lambda_{\uparrow} \approx \lambda_{\downarrow} \approx \lambda_{\uparrow\!\downarrow} \approx 1/4$.
	In this limit, all individual sites become uniformely coupled to the rest of the chain, so that 
	the entropy reaches its maximum. In the presence of coupling, the 
	competition between the scales $t$ and $U$ results in a complex ground-state, whose components 
	are formed by spin configurations with non-trivial occupation probabilities $\lambda_0 \neq \lambda_\up \neq \lambda_\down \neq \lambda_{\updown}$. The sensitivity of $\ket{\Psi_0}$  to the Coulomb repulsion  
	manifests in the measure of the average single-site entanglement, 
	as $\expectv{S_j}$ decreases almost ballistic within the range $1 \leq U/t \leq 10^2$.
	Outside this region and, in particular, for very large coupling, 
	empty and double occupations vanish $\lambda_{0}, \lambda_{\uparrow\!\downarrow} \rightarrow 0$. 
	The entropy reaches half of its maximum, as the components of $\ket{\Psi_0}$ formed by single occupied sites 
	contributes equally $\lambda_{\up} \approx \lambda_{\down} \rightarrow 1/2$. We note that this corresponds to the maximum entropy of the halved Hilbert space.

	We can now examine how far from the thermodynamical limit are these quantities 
	in the case of finite chains under twisted boundary conditions. We vary the torsion of $\pi$
	around the special phases $\Theta_{\text{odd}}^*$ or $\Theta^*_{\text{even}}$ 
	under which the system possesses particle-hole and translation symmetries.
	In order to extend the previous analysis about the correspondence between the  
	the phases of the Hubbard Hamiltonian and its coupling regimes, we keep the Coulomb repulsion 
	within the range $10^{-2} < U/t < 10^{3}$. 
	
	Initially, we consider the deviations in the per-site ground-state energy from the  
	thermodynamical limit of the one-dimensional Hubbard Hamiltonian at half-filling, analysing the 
	dependences on the coupling $U/t$ and the torsion $\Theta$. 
	Here, we argue that working with the absolute difference instead of the percentual deviation 
	is more convenient in the comparison of the ground-state energy because it avoid numerical 
	divergences in the limits ($U \gg 10t$) where $\epsilon_0$ 
	vanishes . 
	Explicitly, $\Delta \epsilon_0$ is calculated as follows 
	\begin{align}
		\label{eq:absolute_dev}
		\Delta \epsilon_0 (L, n=1, U, \Theta) = |\epsilon_0(L, n=1, U, \Theta) - \epsilon_0^{BA}(n=1,U)|.
	\end{align}
	
\begin{figure*}[htb!]
	
	\begin{center}
		\includegraphics[]{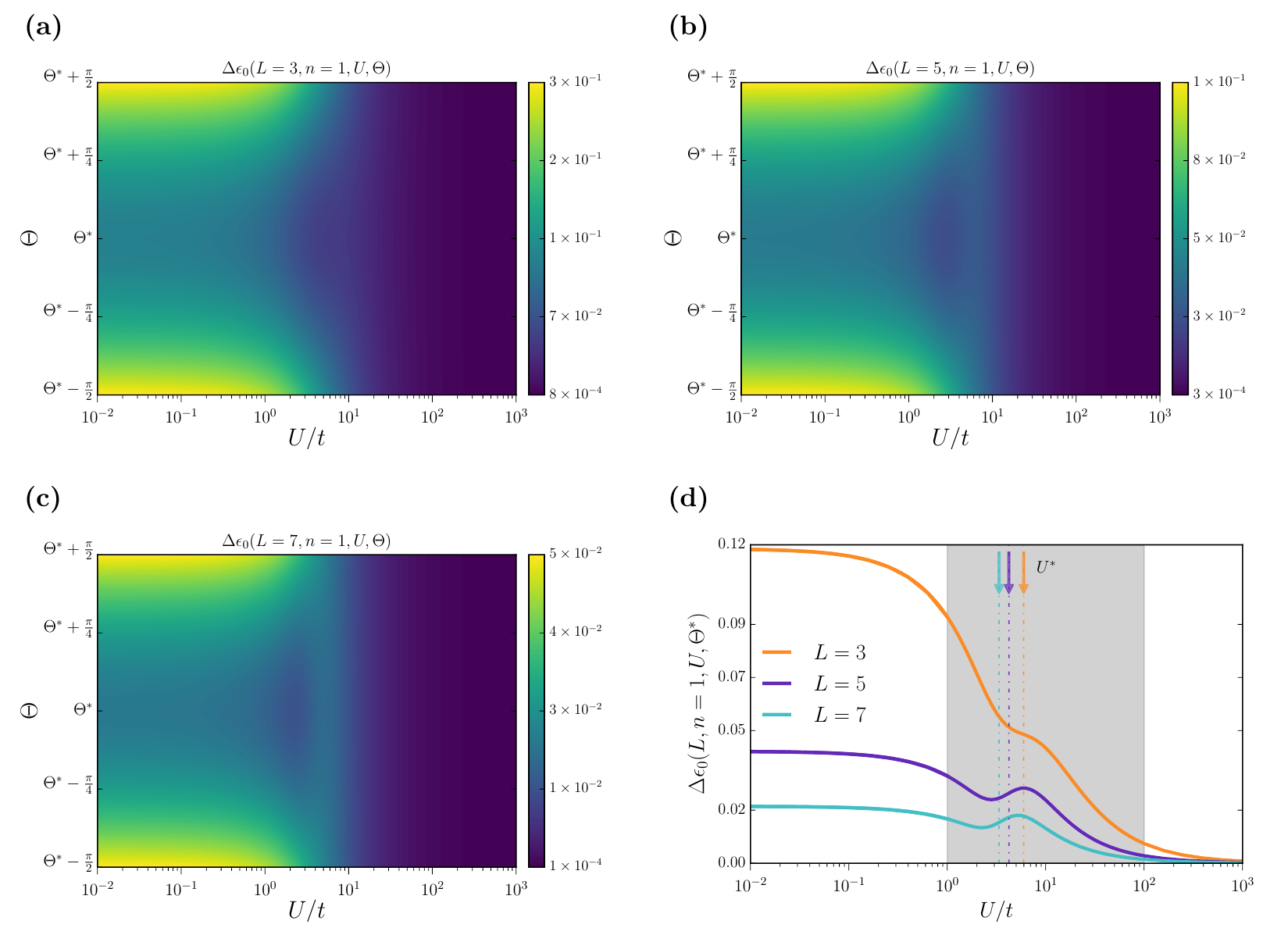}
	\end{center}
	
	\caption{\label{fig:E_gs_odd} Deviations in the per-site ground-state density energies \NEW{ $\epsilon_0(L, n=1, U, \Theta)$} 
	of finite Hubbard chains from 
	the thermodynamical limit \NEW{$\epsilon_0^{BA}(n=1,U)$ calculated from the Bethe Ansatz. }
	Panels \textbf{(a)}, \textbf{(b)} and \textbf{(c)} 
	show $\Delta \epsilon_0(L, n=1, U, \Theta)$ as a function of the torsion $\Theta$ and 
	the coupling $U / t$ for lattices with $L=3,5$ and $7$, respectively.
	Panel \textbf{(d)} displays \NEW{$\Delta \epsilon_0(L, n=1,U,\Theta^*)$} under the special torsion $\Theta^*_{\text{odd}} = \pi/2$ 
	and highlight the couplings $U^*/t$ where the deviations present an inflection point. 
	The highest deviations from the thermodynamical limit $L \rightarrow \infty$
	are found in the non-interacting and intermediate coupling regimes ($U/t < 10$).}
\end{figure*}

\begin{figure*}[htb!]

	\begin{center}
		\includegraphics[]{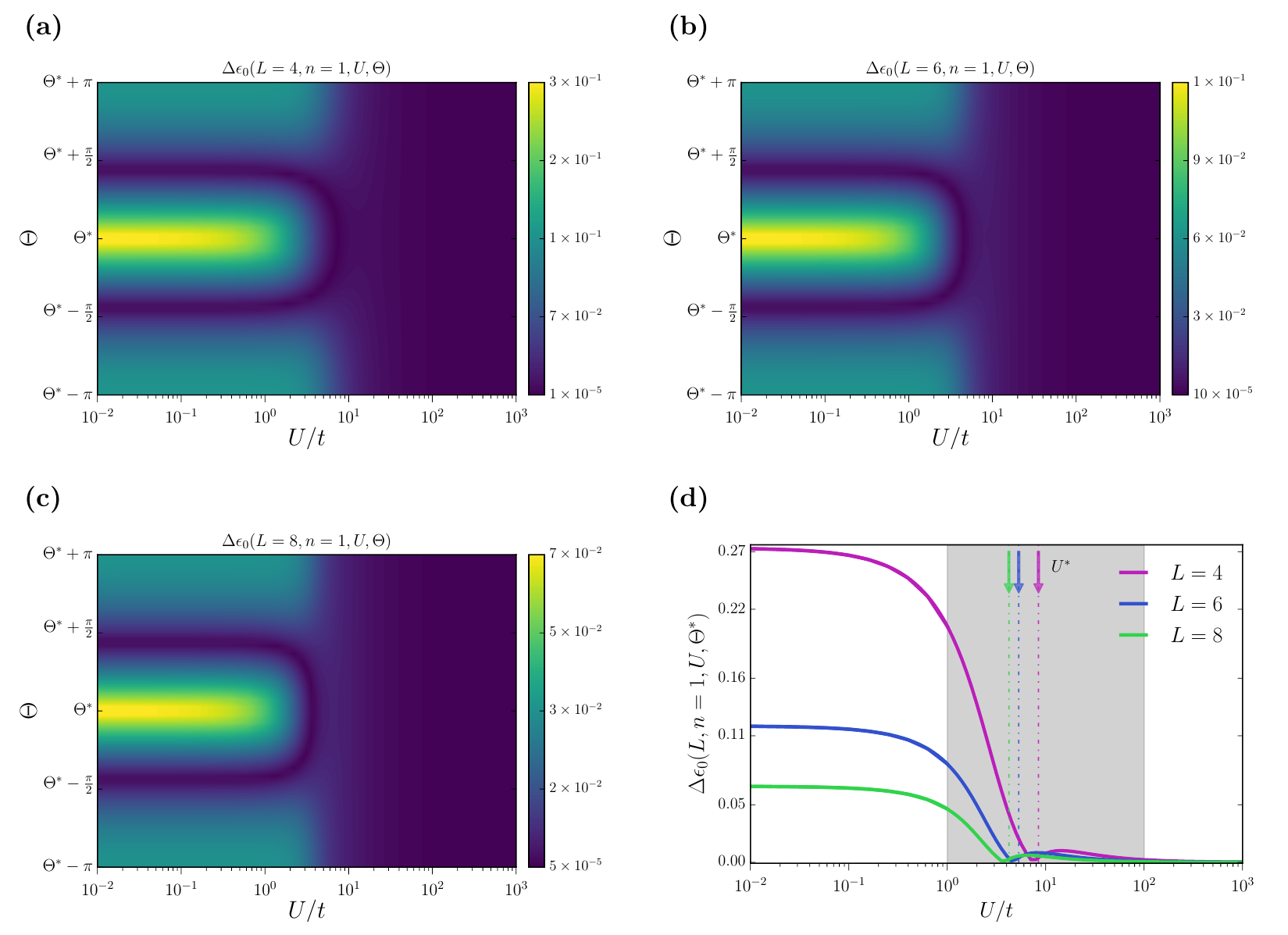}
	\end{center}
	
	\caption{\label{fig:E_gs_even} 
	Deviations in the per-site ground-state density energies \NEW{$\epsilon_0(L, n=1, U, \Theta)$}
	of finite Hubbard chains from 
	the thermodynamical limit \NEW{$\epsilon_0^{BA}(n=1,U)$ calculated from the Bethe Ansatz. }
	Panels \textbf{(a)}, \textbf{(b)} and \textbf{(c)} 
	show $\Delta \epsilon_0(L, n=1, U, \Theta)$ as a function of the torsion $\Theta$ and 
	the coupling $U / t$ for lattices with $L=4,6$ and $8$, respectively.
	Panel \textbf{(d)} displays \NEW{$\Delta \epsilon_0(L, n=1, U, \Theta^*)$} under the special torsion $\Theta^*_{\text{even}} = \pi$. 
	The highest deviations from the thermodynamical limit $L \rightarrow \infty$ 
	are found in the non-interacting and intermediate coupling regimes ($U/t < 10$).
	}

\end{figure*}

	Figures \ref{fig:E_gs_odd} and \ref{fig:E_gs_even} show 
	$\Delta \epsilon_0$ as a function of $U/t$ and $\Theta$ for 
	odd and even number of lattice sites, respectively.
	The dependence of $\Delta \epsilon_0$ on $\Theta$ is only appreciable in low and intermediate coupling regimes, 
	where we identify a periodic behavior which differ among chains with odd and even number of sites. 
	\NEW{
	The difference between $L$ odd and $L$ even can be understood easily by inspecting the 
	non-interacting limit ($U = 0$), for which the deviations are maximized. For $L$ even, the Hamiltonian remains invariant under inversion. Under the special torsion $\Theta^*$ for $U=0$ and even $L$, several of the single-particle levels with nonzero momentum are degenerate. For $L$ odd, the Hamiltonian breaks inversion symmetry and the single-particle levels are not degenerate. As the degeneracy of levels for $L$ even leads to a relatively poor representation of the thermodynamical limit, it follows that $\Theta^*$ preserving particle-hole symmetry maximizes the deviation from the Bethe Ansatz solution for $L \rightarrow \infty$. By contrast, in chains with odd number of sites (for which the single-particle levels are non-degenerate), the special condition minimizes the deviation. This analysis can be extended for the interacting Hubbard Hamiltonian. 
	Moreover, if we consider $\Theta$ varying from $0$ to $L\pi$, we will observe $L$ minima for both even and odd chains, 
	their position being $(2n +1) \pi / 2$ $n=1,...,L$. The maxima, occur in the mid points of the minima and differ between even and odd chains. 
	Inspecting panels \textbf{(a)}-\textbf{(c)} of Fig. \ref{fig:E_gs_odd}, we observe two maxima in the deviations at points 
	$\Theta^* \pm \pi / 2$ for $L=3,5,7$. Panels \textbf{(a)}-\textbf{(c)} of Fig. \ref{fig:E_gs_even} reveal a different structure: the highest deviation from the Bethe Anstaz occur exactly at the special torsion $\Theta^*$, and two local maxima with a smaller amplitude is found at the points $\Theta^* \pm \pi$. This corresponds to a different periodicity around the 
	special torsion $\Theta^*$: for $L$ odd, the behavior repeats around $\Theta^* \pm \pi/2$, while for 
	even $L$, the periodicity of properties occur around $\Theta^* \pm \pi$.}
	
	Comparison of panels \textbf{(a)}-\textbf{(c)} in figures \ref{fig:E_gs_odd} and \ref{fig:E_gs_even}
	within the coupling region limited by $U/t < 10$, indicates the lowest deviations in the energy 
	for $L=3,5$ and $7$ occurying exactly at $\Theta^*_{\text{odd}}$, 
	whereas for $L=4,6$ and $8$, $\Delta \epsilon_0$ reaches its maximum value for $\Theta^*_{\text{even}}$. 
	Clearly, increasing $L$ ensures convergence to the 
	thermodynamical limit. Following panels \textbf{(a)} to \textbf{(c)} in figure \ref{fig:E_gs_odd}, we observe 
	the highest deviations decrease from $\Delta \epsilon_0 \approx 0.3$ for $L=3$ to 
	one order below  $\Delta \epsilon_0 \approx 0.05$ for $L=7$.
	In the case of $L$ even, shown in \ref{fig:E_gs_even}\textbf{(a)}-\textbf{(c)}, we note that 
	the upper limit of $\Delta \epsilon_0$ is of the same order of those found in $L-1$,
	with the correspondence $L=3$ and $L=4$, $L=5$ and $L = 6$, and $L=7$ and $L=8$. 
	When the system approaches the N\`eel state, the lowest absolute differences in energy are 
	$\Delta \epsilon_0 \approx 10^{-4}$ for $L$ up to $7$ and $\leq 5 \times 10^{-5}$ for $L=8$.

	The case in which the special torsion $\Theta^*$ ensures particle-hole symmetry is 
	presented in panel \textbf{(d)} of figures \ref{fig:E_gs_odd} and \ref{fig:E_gs_even}. Under $\Theta^*$, 
	deviations from the thermodynamical limit are nearly constant for $U/t < 1$, and depict a rapid decreasing 
	up to $U/t < 100$,  
	when the system becomes antiferromagnetic. 
	The shaded region distinguishes the limits of couplings for which the system is away from either the 
	single-particle and the N\`eel states.
	For both odd and even chains, $\Delta \epsilon_0(U, \Theta^*)$ depicts a local minimum followed by 
	a local maximum. We observe the inflection points occurying at different positions $U^*(L)$ in the axis $U/t$, 
	indicating the scaling of $\epsilon_0(L, \Theta^*)$ , as for example, $U^*(L=7) < U^*(L=5) < U^*(L=3)$, and similarly, with $L$ even.

	The analysis of the transition from the non-interacting ($U = 0$) to the N\`eel insulating phase ($U/t\rightarrow \infty$) - 
	shaded region in panel \textbf{(d)} in figures \ref{fig:E_gs_odd} and \ref{fig:E_gs_even}  -  
	can be better understood in terms of the average single-site entanglement $\expectv{S_j}$, 
	which has been recently proposed as a \NEW{witness of quantum phase transition 
	\cite{Osterloh-Nat.416.608O,Wu-PRA.74.052335,Vidal-PRA.69.022107,Larsson-PRA.73.042320,Amico-RMP.80.517}. }
	An important observation concerns the homogeinety of single-site entanglement along the chain, which is highly 
	sensible to closed or open boundary conditions. As discussed in Ref. \cite{Zawadzki-BJP.47.5}, 
	local densities and magnetizations vary from site to site under open boundary conditions.
	Under twisted boundary conditions, a special case of closed boundary condidions, 
	the densities are uniform and independent of $U/t$ and so does $S_j = \expectv{S_j}$. 
	Nevertheless, the strength of $U/t$ modifies the inner structure 
	of the ground-state wave-function and this dependence must be reflected in correlation measurements, 
	such as the entanglement. In that sense, our proposal to examine the effects of the twisted boundary 
	condition in the average single-site entanglement can help 
	to identify degrees of freedom contributing to the ground-state. Also, analysing the deviations 
	from the infinite system can provide a deep understanding of 
	the role of symmetries in connecting effective correlation lenghts 
	to produce states and phases of the thermodynamical limit.

\begin{figure}[htb!]

	\begin{center}
		\includegraphics[]{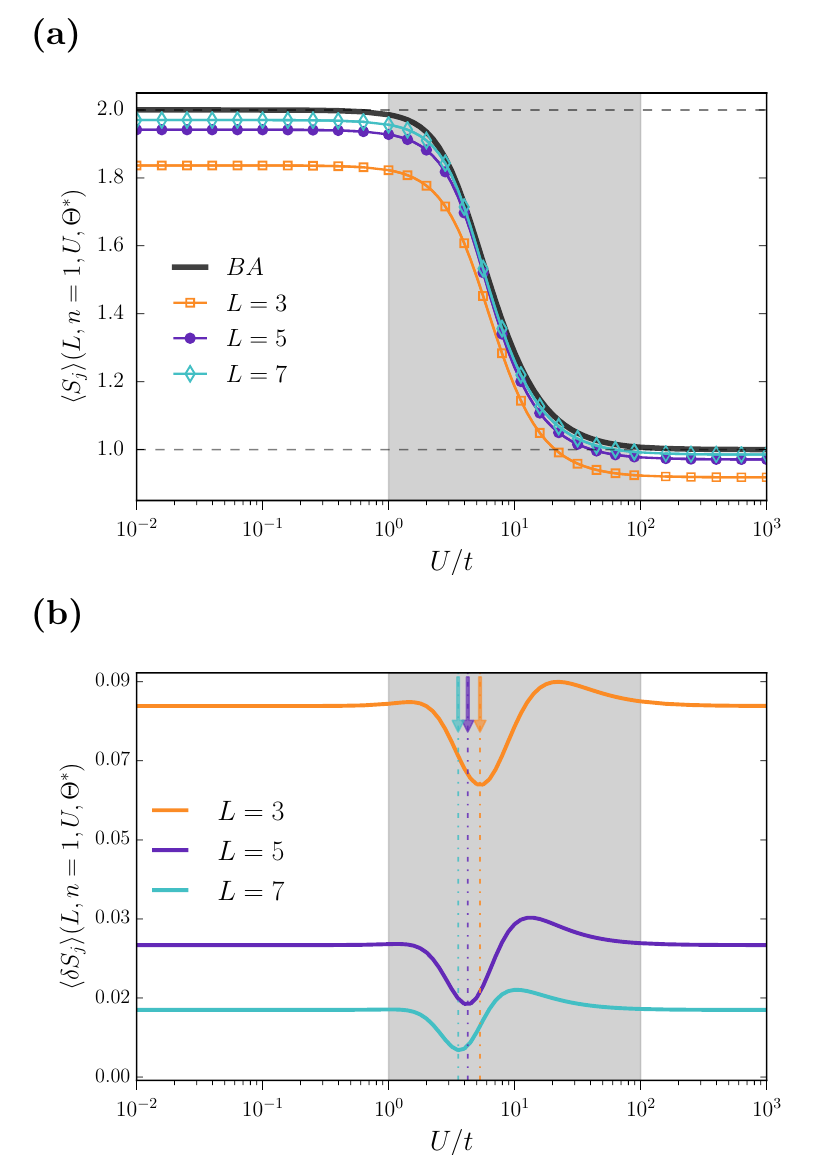}
	\end{center}

	\caption{\label{fig:SSem_odd} Mean single site entanglement $\expectv{S_j}$ 
	as a function of the coupling $U/t$ for chains with odd number of sites under torsion $\Theta^*$. 
	\textbf{(a)} The estimate for $\langle S_j \rangle$ in the limit $L \rightarrow \infty$ obtained via Bethe Ansatz 
	is represented in black solid lines and colored marked curves show 
	$\expectv{S_j}(L, n=1, U, \Theta^*)$ for chains of odd number of sites $L = 3,5,7$.
	\textbf{(b)} Percentual difference $\expectv{\delta S_j}(L,n=1,U,\Theta^*)$ from the thermodynamical limit as a function of $U/t$.
	Colored arrows on the top of the right panel indicates the scaled couplings $U/t$ for which the relative deviations $\expectv{\delta S_j}$ are 
	minimum. 
	}

\end{figure}

\begin{figure}[htb!]

	\begin{center}
		\includegraphics[]{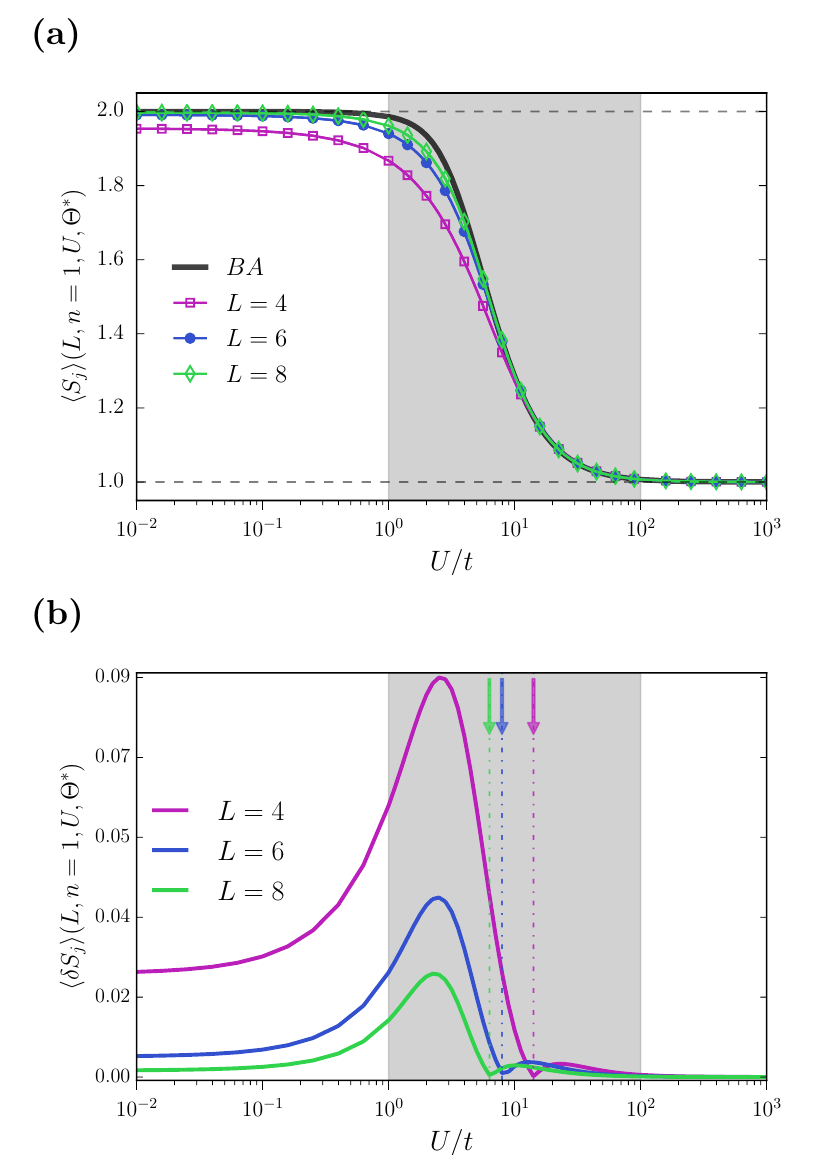}
	\end{center}
	
	\caption{\label{fig:SSem_even} Mean single site entanglement $\expectv{S_j}$ 
	as a function of the coupling $U/t$ for chains with even number of sites under torsion $\Theta^*$. 
	\textbf{(a)} The estimate for $\langle S_j \rangle$ in the limit $L \rightarrow \infty$ obtained via Bethe Ansatz 
	is represented in black solid lines and colored marked curves show 
	$\expectv{S_j}(L, n=1, U, \Theta^*)$ for chains of odd number of sites $L = 4,6,8$.
	\textbf{(b)} Percentual difference $\expectv{\delta S_j}(L,n=1,U,\Theta^*)$ from the thermodynamical limit as a function of $U/t$.
	Colored arrows on the top of the right panel indicates the scaled couplings $U/t$ for which the relative deviations $\expectv{\delta S_j}$ are 
	minimum.
	}

\end{figure}

	Our results for $\expectv{S_j}(L, n=1, U, \Theta^*)$ as a function of the coupling $U/t$ for finite chains under $\Theta^*$ 
	are presented in figures \ref{fig:SSem_odd} and \ref{fig:SSem_even}. In panels \textbf{(a)},  
	the entanglement entropy for the infinite Hubbard model - calculated from eq. \refeq{eq:SSem_BA} - 
	is represented by a solid black line, whereas 
	$\expectv{S_j}(L, n=1, U, \Theta^*)$ calculated for $L=3,4,5,6,7,8$ is shown in colored lines and markers.
	We note that $\expectv{S_j}$ is lower than 
	its value in the infinite Hubbard model for $L$ odd in all coupling regimes, while 
	for even $L$ it stays within the limits $1 \leq S_j(L\rightarrow \infty) \leq 2$. The reason for that is that at half-filling, 
	chains with odd number of sites are magnetized with total spin $S = 1/2$, so that $\lambda_{\up} > \lambda_{\down}$ for $U \gg 1$. 
	In the extreme, $U/t \rightarrow \infty$ the N\`eel state is described by two components with weights 
	corresponding to $\lambda_\up$ and $\lambda_\down$. For closed chains with odd $L$, the antiferromagnetic state 
	arising in the limit $U/t \gg 100$ offers an example of magnetic frustration, 
	absent for even $L$ as spins $\up$'s match consecutive $\down$'s.

	For the deviations in the average single-site entanglement, we work with the percentual difference $\expectv{\delta S_j}$ between 
	the Bethe Ansatz estimate $\expectv{S_j}^{BA}(n=1,U)$ and the calculated $\expectv{S_j}(L, n=1,U,\Theta)$ for finite chains. 
	Explicitily, 
	\begin{align}
		\label{eq:percent_diff_S}
		\expectv{\delta S_j} (L, n=1, U, \Theta) & = 
		\frac{|\expectv{S_j}(L, n=1,U,\Theta) - \expectv{S_j}^{BA}|}{\expectv{S_j}^{BA}}
		.
	\end{align}

	Panel \textbf{(b)} of figures \ref{fig:SSem_odd} 
	and \ref{fig:SSem_even} display $\expectv{\delta S_j}$ as a function of $U/t$ 
	for chains with  $L = 3,5,7$ and $L = 4,6,8$ in the case where the torsion is $\Theta^*$.
	Similarly to the ground-state energy, chains with odd and even number of sites present opposite trends under $\Theta^*$	
	The behavior of $\expectv{\delta S_j} (L, n=1, U, \Theta)$ is particularly interesting within the shaded region ($1 \leq U/t \leq 100$).
	For $L$ odd, the deviation $\expectv{\delta S_j}$ is nearly constant for $U/t \leq 1$ and $U / t \geq 10$. The cases $L=4,6,8$ 
	depict a different trend, the differences in $\expectv{S_j}$ start increasing in couplings of one order below 
	 those with successive $L$ even, 
	reaching a maximum value for all even sizes around $U/t \approx 2.5$ and smoothly decreasing to valleys in 
	$U/t \approx 6.5, 8.5 $ and $14.5$ for $L=4,6$ and $8$, respectively. The couplings $U/t$ 
	for which $\expectv{\delta S_j}$ is minimum are marked in colored arrows on the top of the panels 
	\ref{fig:SSem_odd} \textbf{(b)} and \ref{fig:SSem_even} \textbf{(b)}. Following the increasing in the chain size, 
	we observe $U/t$ to decrease, suggesting not only a scale property, but also the existence of 
	a critical coupling for which a finite system with size $L$ is able to reproduce with arbitrarily good precision the 
	correlations of the thermodynamical limit.
	A deep understanding of such property requires 
	further examination; we suggest to investigate other correlation measures, such as 
	the spin correlations and block-block entanglement.

\begin{figure}[htb!]

	\begin{center}
		\includegraphics[]{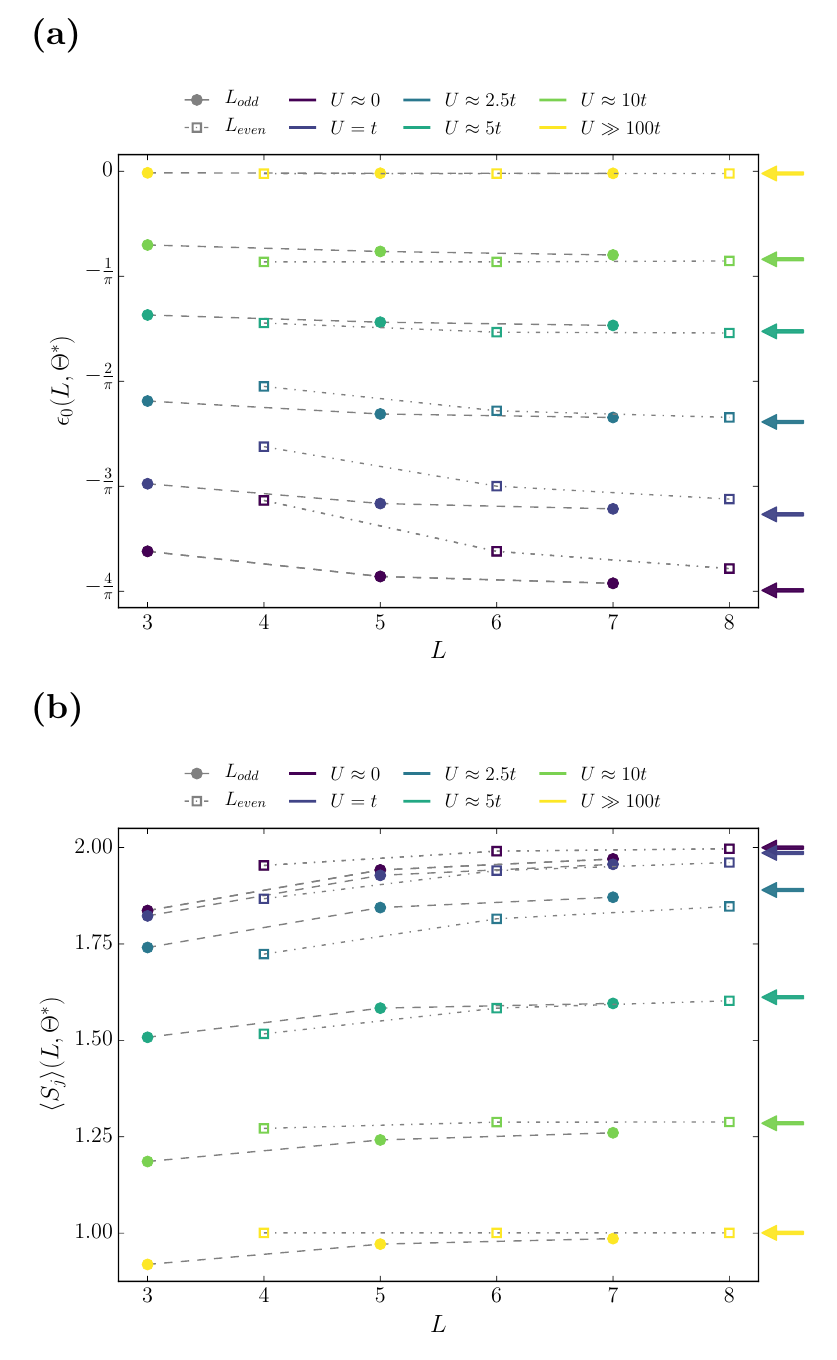}
	\end{center}
	
	\caption{\label{fig:scaling} Scaling of the ground-state energy \textbf{(a)} and average single-site entanglement \textbf{(b)} under twisted boundary 
	conditions with torsion $\Theta^*$ for various coupling regimes. Empty squares represent $L$ even and filled circles correspond 
	to $L$ odd.  \NEW{The different scaling trend followed by $L$ even (dot-dashed lines) and odd (dashed lines) is understood in terms of the degeneracy of spin configurations contributing to the ground-state in the case of $L$ even under $\Theta^*$. The degeneracy under particle-hole symmetry for $L$ even reduces the number of effective states needed to represent the thermodynamical limit.}
	Values of $\epsilon_0^{BA}(n=1,U)$ and $\expectv{S_j}^{BA}(n=1,U)$ in the thermodynamical limit ($L \rightarrow \infty$)
	are indicated by colored arrows on the right side of the axis.
	.}

\end{figure}

	Finally, we analyse the scaling on the ground-state and the average single-site entanglement under $\Theta^*$.
	Figure \ref{fig:scaling} shows $\epsilon_0(L, n=1, U, \Theta^*)$ \textbf{(a)} and $\expectv{S_j}(L, n=1, U, \Theta^*)$ 
	\textbf{(b)} as a function of $L$ for some values of $U/t$ ranging from the free (dark blue) to the strongly coupling (yellow) regime. 
	Colored arrows on the right side of the panels indicate the values of $\epsilon_0(U, n=1)$ and $\expectv{S_j}(U, n=1)$ for 
	$L \rightarrow \infty$. Colored circles and empty squares identify odd and even chains, respectively. Comparing 
	chains with odd and even number of sites, 
	we note the first perform better in low and intermediate coupling regimes. For $U/t > 1$, they become comparable, and for $U/t > 10$, there is an inversion, as values for even $L$ 
	are closer to the Bethe Ansatz. The same trend is observed for both ground-state energy and single-site entanglement.

\section{\label{sec:conclusion} Conclusions}

	We have discussed the one-dimensional finite Hubbard Hamiltonian under twisted boundary conditions and 
	examined two important invariances present in the infinite model, namely, particle-hole symmetry and momentum conservation. 
	We have derived the special torsion phase which restores these symmetries in finite Hubbard chains 
	by means of local twisted hoppings with phases of $\Theta^*_{\text{odd}} = \pi /2$ and $\Theta^*_{\text{even}} = \pi$. 
	We have presented exact numerical results for the ground-state energy of half-filled chains 
	as a function of the torsion $\Theta$ and the coupling $U/t$, investigating 
	how far from the thermodynamical limit these quantities are for chains with size $L=3,4,5,6,7,8$ 
	under the special torsion. We show that, ensuring particle-hole and translation symmetry by 
	fixing $\Theta^*$, the \NEW{deviations in the per-site ground-state energy
	of lattices of few sites ($L=7$ or $L=8$) from the Bethe Ansatz calculation 
	for $L \rightarrow \infty$ are maximum in small and intermediate coupling regimes}, whereas reproduces quite well the 
	insulating phase of the infinite Hubbard model.
	The analysis of the average single-site entanglement completed our analysis of the 
	phase transition in finite Hubbard lattices and of its scaling behavior. We have identified couplings 
	for which finite lattices enter in the N\`eel antiferromagnetic insulating phase.  Finally, we discussed the differences 
	between chains with even and odd number of sites. 
	Our findings provide new insights into the understanding of scaling laws in phase transitions 
	occurying in finite systems. In particular, examining the role of symmetries 
	in finite chains and their correspondence with the thermodynamical limit can help us to
	\NEW{identify the quantum states yielding the
most important contributions to the $L\rightarrow \infty$
limit. Special attention to such states may
help us to define novel renormalization-group
transformations. }
	Moreover, the understanding of symmetries presevation in few particle systems has a 
	practical importance for  
	quantum technologies, as it can guide the development of protocols for manipulating properties in 
	qubits systems.

\begin{acknowledgements}

LNO acknowledges FAPESP (Fellowship grant no. 12/02702-0) and CNPq (grants no. 312658/2013-3) for financial support. 
KZ aknowledges support from CNPq (PhD Scholarship grant no. 140703/2014-4) and CAPES 
(PDSE grant no. 88881.135185/2016-01). ID acknowledges 
support from the Royal Society through the Newton Advanced Fellowship scheme (grant no. NA140436) and CNPq through 
the PVE scheme (grant no. 401414/2014-0).

\end{acknowledgements}

\newpage

\bibliography{hubbard_entanglement}

\newpage
\appendix

\section{\label{apx:Sj_Theta} Single-site entanglement deviations as a function of $\Theta$ and $U/t$}

	In section \ref{sec:results}, we presented results (Figs. \ref{fig:SSem_odd} and \ref{fig:SSem_even}) for the deviations $\delta S_j$ from the thermodynamical limit in the single-site entanglement of finite Hubbard chains under the special torsion $\Theta^*$. Here, we present the results for $\delta S_j$ as a function of the torsion $\Theta$ and the coupling $U/t$.
	
\begin{figure*}[htb!]

	\begin{center}
		\includegraphics[]{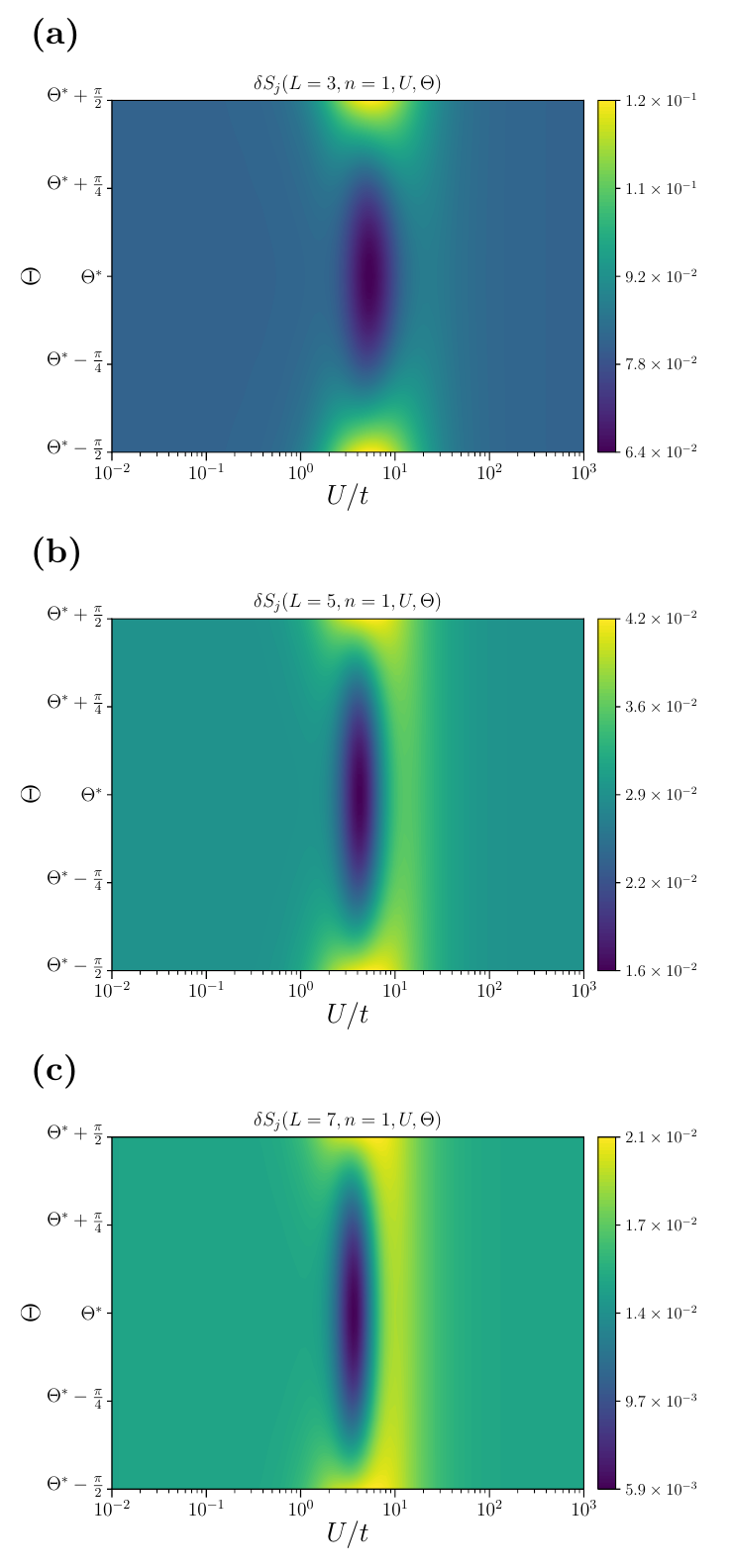}
	\end{center}
	
	\caption{\label{fig:SSe_gs_odd} 
	Deviations in the average single-site entanglement $\expectv{S_j}(L,n=1,U, \Theta)$
	of finite Hubbard chains from 
	the thermodynamical limit $\expectv{S_j}^{BA}(n=1,U)$ calculated from the Bethe Ansatz. 
	Panels \textbf{(a)}, \textbf{(b)} and \textbf{(c)} 
	show $\delta S_j(L, n=1, U, \Theta)$ as a function of the torsion $\Theta$ and 
	the coupling $U / t$ for lattices with $L=3,5$ and $7$, respectively.
	The lowest deviations from the thermodynamical limit $L \rightarrow \infty$ 
	are found in the intermediate coupling regimes ($1<U/t < 10$), the minimum deviation occurying under $\Theta^*$.
	}

\end{figure*}

	Figure \ref{fig:SSe_gs_odd} shows the deviations in the average single-site entanglement for chains with an odd number of sites $L=3,5$ and $7$. Similarly to the periodic ehavior with respect to $\Theta$ observed in the plots for the ground-state energy $\Delta \epsilon_0(L,n=1,U,\Theta)$, the deviations $\delta S_j (L,n=1,U,\Theta)$ for odd $L$are minima under the special torsion $\Theta^*$ and maxima at the points $\Theta^* \pm \pi/2$. For odd $L$, the deviations are constant in small and strong coupling regimes; for intermediate coupling regimes $1 < U/t < 10$, a complex structure arises.

\begin{figure*}[htb!]

	\begin{center}
		\includegraphics[]{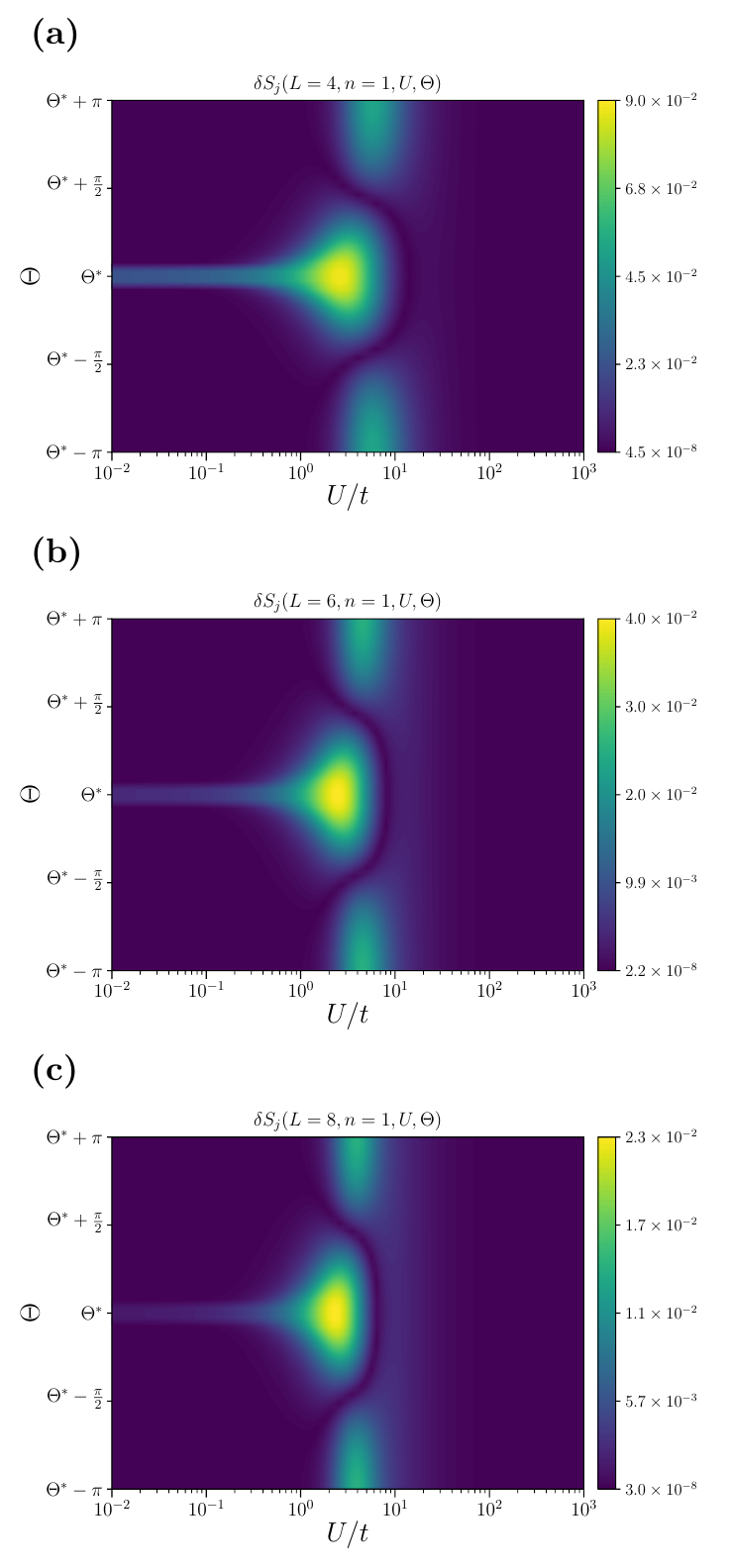}
	\end{center}
	
	\caption{\label{fig:SSe_gs_even} 
	Deviations in the average single-site entanglement $\expectv{S_j}(L,n=1,U, \Theta)$
	of finite Hubbard chains from 
	the thermodynamical limit  $\expectv{S_j}^{BA}(n=1,U)$ calculated from the Bethe Ansatz. 
	Panels \textbf{(a)}, \textbf{(b)} and \textbf{(c)} 
	show $\delta S_j(L, n=1, U, \Theta)$ as a function of the torsion $\Theta$ and 
	the coupling $U / t$ for lattices with $L=4,6$ and $8$, respectively.
	The highest deviations from the thermodynamical limit $L \rightarrow \infty$ 
	are found in the intermediate coupling regimes ($1<U/t < 10$), the maximum deviation occurying under $\Theta^*$.
	}

\end{figure*}

	The percentual deviations from the Bethe Ansatz in the case of chains with even number of sites is shown in figure \ref{fig:SSe_gs_even}. The deviation is below $10^{-8}$ in small and strong coupling regimes, being amplified for couplings $1 < U/t < 10$ specially at the torsion $\Theta^*$ ensuring particle-hole symmetry.

\clearpage	
\newpage

\section{\label{apx:trimer} Case study: trimer}	
	For instance, consider a trimer $L = 3$ with $\mu=0$. 	For open and periodic boundary conditions, 
	Hamiltonian of eq. \refeq{eq:Hubbard_BC} can be written as 
\begin{align}
	\label{eq:H_trimer_OBC}
	\hat{H}_{OBC}(L=3) & =  - t (\Cop_{1 \sigma}^\dagger \Cop_{2 \sigma} + \Cop_{2 \sigma}^\dagger \Cop_{1 \sigma}) 
				 - t (\Cop_{2\sigma}^\dagger \Cop_{3\sigma} + \Cop_{3\sigma}^\dagger \Cop_{2\sigma})
			\nonumber \\ & \hphantom{=}	  
			+ U \Nop_{1\up} \Nop_{1 \down} + U \Nop_{2\up} \Nop_{2 \down} + + U \Nop_{3\up} \Nop_{3 \down}  
\end{align}	
and
\begin{align}
	\label{eq:H_trimer_PBC}
	\hat{H}_{PBC}(L=3) & =  \hat{H}_{OBC}(L=3) - t (\Cop_{1 \sigma}^\dagger \Cop_{3 \sigma} + \Cop_{3 \sigma}^\dagger \Cop_{1 \sigma}),
\end{align}	
respectively.	
	
	For twisted boundary conditions with twist phase $\theta$ we replace $\Cdop_\ell \rightarrow e^{i\theta \ell} \Cdop_\ell$ to write  
	the Hamiltonian \refeq{eq:Hubbard_BC} as 
\begin{align}
	\label{eq:H_trimer_TBC}
	\hat{H}_{TBC}(L=3) & = 
	-t e^{-i\theta} 
	(
	\Cdop_{1} \Cop_{2} + \Cdop_{2} \Cop_{3} + \Cdop_{3} \Cop_{1}
	)
	\nonumber \\& \hphantom{=}
	- t e^{i\theta} 
	(
	\Cdop_{2} \Cop_{1} + \Cdop_{3} \Cop_{2} + \Cdop_{1} \Cop_{3}  
	)
	\nonumber \\& \hphantom{=}
	+ U \Nop_{1\up} \Nop_{1 \down} + U \Nop_{2\up} \Nop_{2 \down} + + U \Nop_{3\up} \Nop_{3 \down} 
	.
\end{align}

	At half-filling the basis set comprises $18$ states, $9$ associated 
	with $z$ component of spin $S_z = -1/2$  and $9$ with $S_z = +1/2$, which are degenerate. They are:
	\begin{align}
		\label{eq:basis_trimer_half_filling}
		\ket{L=3, S=1/2, S_z=1/2,b = 1} & = \upOp{2} \updownOp{3} \ket{0}
		\nonumber \\
		\ket{L=3, S=1/2, S_z=1/2,b = 2} & =\updownOp{2} \upOp{3} \ket{0}
		\nonumber \\
		\ket{L=3, S=1/2, S_z=1/2,b = 3} & =\downOp{1} \upOp{2} \upOp{3} \ket{0} 
		\nonumber \\
		\ket{L=3, S=1/2, S_z=1/2,b = 4} & = \upOp{1} \updownOp{3} \ket{0}
		\nonumber \\
		\ket{L=3, S=1/2, S_z=1/2,b = 5} & = \upOp{1} \downOp{2} \upOp{3} \ket{0}
		\nonumber \\
		\ket{L=3, S=1/2, S_z=1/2,b = 6} & = \upOp{1} \upOp{2} \downOp{3} \ket{0}
		\nonumber \\
		\ket{L=3, S=1/2, S_z=1/2,b = 7} & = \upOp{1} \updownOp{2} \ket{0}
		\nonumber \\
		\ket{L=3, S=1/2, S_z=1/2,b = 8} & =\updownOp{1} \upOp{3} \ket{0}
		\nonumber \\
		\ket{L=3, S=1/2, S_z=1/2,b = 9} & = \updownOp{1} \upOp{2} \ket{0}		
		.
	\end{align}

	Projecting the operator $\hat{H}$ in eq. \refeq{eq:H_trimer_TBC} into the basis defined by eq. \refeq{eq:basis_trimer_half_filling}, 
	we obtain the corresponding matrix Hamiltonian 
\begin{equation}
\label{eq:Htrimer_TBC_Sundef}
H_{L=3} (\theta)= t
\left(\begin{array}{ccccccccc}
0 & 0 & 0 & -e^{-i\theta} & 0 & e^{i\theta} & 0 & e^{-i\theta} & -e^{i\theta}\\
0 & 0 & 0 & 0 & e^{i\theta} & -e^{i\theta} & e^{-i\theta} & -e^{-i\theta} & 0\\
0 & 0 & 0 & e^{-i\theta} & -e^{i\theta} & 0 & -e^{-i\theta} & 0 & e^{i\theta}\\
-e^{i\theta} & 0 & e^{i\theta} & \tilde{U} & e^{-i\theta} & -e^{-i\theta} & 0 & 0 & 0\\
0 & e^{-i\theta} & -e^{-i\theta} & e^{i\theta} & \tilde{U} & 0 & 0 & -e^{i\theta} & 0\\
e^{-i\theta} & -e^{-i\theta} & 0 & -e^{i\theta} & 0 & \tilde{U} & e^{i\theta} & 0 & 0\\
0 & e^{i\theta} & -e^{i\theta} & 0 & 0 & e^{-i\theta} & \tilde{U} & 0 & -e^{-i\theta}\\
e^{i\theta} & -e^{i\theta} & 0 & 0 & -e^{-i\theta} & 0 & 0 & \tilde{U} & e^{-i\theta}\\
-e^{-i\theta} & 0 & e^{-i\theta} & 0 & 0 & 0 & -e^{i\theta} & e^{i\theta} & \tilde{U}
\end{array}\right),
\end{equation}	
where $\tilde{U} = U / t$.

	Notice that, the periodic boundary condition, can be recovered by choosing $\theta = 0$.
	As discussed in sec. \ref{sec:Hubbard}, $\theta = \pi /2$ ensures particle-hole symmetry. 
		
	In the non-interacting case, $U = 0$ with $\theta = \pi / 2$ the ground-state is a combination of three states 
\begin{align}
\label{eq:GS}
\ket{\Psi_0} & = \frac{1}{\sqrt{3}} \ket{\psi_0}  + \frac{1+ i \sqrt{3}}{2\sqrt{3}} \ket{\psi_1}
	- \frac{1 - i \sqrt{3}}{2\sqrt{3}} \ket{\psi_2},
\end{align}	
where 
\begin{align}
	\label{eq:eigenbasis_trimer}
	\ket{\psi_0} & = \frac{1}{\sqrt{3}} (\ket{\up \, \, \up \, \, \down \, }+  \ket{\up \! \down  \emptys \up} - \ket{\emptys \up \! \down \up })
	\nonumber \\
	\ket{\psi_1} & = \frac{1}{\sqrt{3}} (  - \ket{\up \, \, \down \, \,  \up \, }  + \ket{\up \! \down \, \up \emptys} - \ket{\emptys \up \,  \up \! \down })
	\nonumber \\
	\ket{\psi_2} & = \frac{1}{\sqrt{3}} ( \ket{\down \, \, \up \, \,  \up \, } + \ket{\up \,  \up \! \down  \emptys} - \ket{ \up \emptys \up \! \down} ).
\end{align}
	
	In order to calculate the single-site entanglement, we must compute the reduced density matrix by tracing the degrees of fredoom of two sites, i.e.,
\begin{align}
	\label{eq:reduced_density_matrix_trimer}
	\rho^{k} = \Trace_{\ell \neq k} \ket{\Psi_0} \bra{\Psi_0},
\end{align}	
where $i,j,k$ refers to the sites labels.	

On basis of $\ket{\psi}$'s the density matrix is 
\begin{align}
	\label{eq:rhomat+}
	\rho = \left( \begin{array}{ccc}
	\frac{1}{3} & \frac{1+i\sqrt{3}}{6} & -  \frac{1-i\sqrt{3}}{6} \\
	\frac{1-i\sqrt{3}}{6} & \frac{1}{3} & \frac{1+i\sqrt{3}}{6} \\
	- \frac{1+i\sqrt{3}}{6} & \frac{1-i\sqrt{3}}{6} & \frac{1}{3}
	\end{array}
	\right).
\end{align}

	Let's first consider the trace over sites $0$ and $1$:
\begin{align}
	\label{eq:Tr01}
	\rho_{2} & = \Trace_{0,1} \big[ \rho \big]
	\nonumber \\
	& = \sum_{\sigma_0, \sigma_1} \bra{\sigma_0 \, \sigma_1 } \rho^+ \ket{\sigma_0 \, \sigma_1 },
\end{align}
where $\sigma = \up, \down$.
								
\begin{align}
	\label{eq:Tr01}
	\rho_{2} & =  \Trace_{1,3} \big[ \rho \big] .
\end{align}																									

	The reduced density matrix is diagonal since the products of states of sites $0$ and $1$ with each $\ket{\Psi}$ are
\begin{align}
	\label{eq:products}
	\braket{\sigma_0 \, \sigma_1}{ \psi_0} & = \frac{1}{\sqrt{3}} \Big[  
	\delta_{\sigma_0, \up} \delta_{\sigma_1, \up} \ket{\down} 
	+ \delta_{\sigma_0, \updown} \delta_{\sigma_1, \emptys} \ket{\up}
	- \delta_{\sigma_0, \emptys} \delta_{\sigma_1, \updown} \ket{\up}  
	\Big]
	\nonumber 
	\\
	\braket{\sigma_0 \, \sigma_1}{ \psi_1} & = \frac{1}{\sqrt{3}} \Big[  
	- \delta_{\sigma_0, \up} \delta_{\sigma_1, \down} \ket{\up} 
	+ \delta_{\sigma_0, \updown} \delta_{\sigma_1, \up} \ket{\emptys}
	- \delta_{\sigma_0, \emptys} \delta_{\sigma_1, \up} \ket{\updown}  
	\nonumber 
	\\
	\braket{\sigma_0 \, \sigma_1}{ \psi_2} & = \frac{1}{\sqrt{3}} \Big[  
	\delta_{\sigma_0, \down} \delta_{\sigma_1, \up} \ket{\up} 
	+ \delta_{\sigma_0, \up} \delta_{\sigma_1, \updown} \ket{\emptys}
	- \delta_{\sigma_0, \up} \delta_{\sigma_1, \emptys} \ket{\updown}  	
	\Big],
\end{align}
yielding
\begin{align}
	\label{eq:reduced_Tr01}
	\rho_{2} = \frac{1}{3}\left(
	\begin{array}{cccc}
	|\alpha_1|^2 + |\alpha_2|^2 & 0 & 0 & 0 \\
	0 & |\alpha_0|^2 & 0 & 0 \\
	0 & 0 & 2|\alpha_0|^2 + |\alpha_1|^2 + |\alpha_2|^2 & 0 \\
	0 & 0 & 0 & |\alpha_1|^2 + |\alpha_2|^2
	\end{array}
	\right),
\end{align}		
where 
\begin{align}
	\label{eq:alphas}
	\alpha_0 & = \frac{1}{\sqrt{3}} \rightarrow |\alpha_0|^2 = \frac{1}{3} \nonumber \\
	\alpha_1 & = \frac{1+i\sqrt{3}}{6} \rightarrow |\alpha_2|^2 = \frac{1}{3} \nonumber \\
	\alpha_2 & = \frac{1-i\sqrt{3}}{6} \rightarrow |\alpha_2|^2 = \frac{1}{3}, 
\end{align}				
so that eq. \refeq{eq:reduced_Tr01} is expressed as 
\begin{align}
	\label{eq:reduced_Tr01}
	\rho_{2}  = &  
	\left(
	\begin{array}{cccc}
	\frac{2}{9} & 0 & 0 & 0 \\
	0 & \frac{1}{9} & 0 & 0 \\
	0 & 0 & \frac{4}{9} & 0 \\
	0 & 0 & 0 & \frac{2}{9}
	\end{array}
	\right).
\end{align}	

	The mean single-site entanglement is therefore 
	\begin{align}
		\label{eq:SSetrimer}
		S_2 (U = 0)& = - \frac{1}{9} 
		\left[ 4 \log(\frac{2}{9})
		+ \log(\frac{1}{9})
		+ 4 \log(\frac{4}{9})
		 \right]
		 \nonumber \\
		 & = \approx 1.8365.
	\end{align}

	The Mott-insulating phase ($U/t \rightarrow \infty$) of the Hubbard trimer 
	is the other limit in which analytical calculations are straightforward.
	The high price for double occupation reduces the basis set in eq. \refeq{eq:basis_trimer_half_filling}  
	to only three components $\ket{\down \up \up }$, $\ket{\down \up \down}$ and $\ket{\up \up \down}$.
	The probabilities to have empty ($\lambda_{\_}$) and double occupied ($\lambda_{\updown}$) sites vanishes. Once we fixed 
	the magnetization of the system to be $m = \frac{1}{3}$, $\lambda_\up = \frac{2}{3}$ and 
	$\lambda_\down = \frac{1}{3}$, so that the single-site entanglement is 
	\begin{align}
		\label{eq:SSetrimer}
		S_2 (U \rightarrow \infty)& = - \frac{1}{3} 
		\left[ 2 \log\left( \frac{2}{3} \right) + \log\left( \frac{1}{3} \right)
		 \right]
		 \nonumber \\
		 & = \approx 0.9183.
	\end{align}

\end{document}